\documentclass[sts]{imsart}

\RequirePackage{amsthm,amsmath,amsfonts,amssymb}
\RequirePackage[numbers]{natbib}
\usepackage{diagbox}
\usepackage{booktabs}
\usepackage{graphicx}
\usepackage[table]{xcolor}
\usepackage{amsmath}
\usepackage{amsfonts}
\usepackage{bm}
\usepackage{hhline}
\usepackage{array}
\usepackage{multirow}

\makeatletter
\let\tablewidth\relax
\makeatother

\usepackage{tabularray}
\UseTblrLibrary{booktabs} 

\usepackage{hyperref}
\usepackage{algorithm}
\usepackage{algorithmic}
\usepackage{listings}
\usepackage[algo2e,ruled,vlined]{algorithm2e}
\usepackage{enumitem}
\usepackage{adjustbox}
\usepackage{hyperref}
\usepackage{pdflscape}
\usepackage{subfig}
\usepackage{mathtools}
\usepackage{float}
\usepackage{placeins}

\startlocaldefs

\endlocaldefs

\begin{document}

\begin{frontmatter}
\title{Leveraging external data in the analysis of randomized controlled trials: a comparative analysis}
\runtitle{Leveraging external data in the analysis of randomized controlled trials}

\begin{aug}
\author[A]{\fnms{Gopal}~\snm{Kotecha*}\ead[label=e1]{gkotecha@g.harvard.edu}},
\author[B]{\fnms{Daniel E.}~\snm{Schwartz*}\ead[label=e2]{dschwartz12@mgh.harvard.edu}},
\author[C]{\fnms{Steffen}~\snm{Ventz}\ead[label=e3]{ventz001@umn.edu}}
\and
\author[D]{\fnms{Lorenzo}~\snm{Trippa}\ead[label=e4]{ltrippa@jimmy.harvard.edu}}
\address[A]{Gopal Kotecha, QLS Advisors, Cambridge, USA.}
\address[B]{Daniel E. Schwartz, Massachusetts General Hospital Biostatistics \printead[presep={\ }]{e2}.}
\address[C]{Steffen Ventz, University of Minnesota School of Public Health, Division of Biostatistics \& Health Data Science.}
\address[D]{Lorenzo Trippa, Harvard T.H. Chan School of Public Health, Department of Biostatistics. * = co-first authors.}

\end{aug}

\begin{abstract}
The use of patient-level information from previous studies, registries, and other external datasets can support the analysis of single-arm and randomized controlled trials to evaluate and test experimental treatments. However, the integration of external data in the analysis of clinical trials can also compromise the scientific validity of the results due to selection bias, study-to-study differences, unmeasured confounding, and other distortion mechanisms. Therefore, leveraging external data in the analysis of a clinical trial requires the use of appropriate methods that can detect, prevent or mitigate the risks of bias and potential distortion mechanisms. We review several methods that allow investigators to leverage external datasets, such as propensity score procedures and random effects modeling. Different methods present distinct trade-offs between risks and efficiencies. We conduct a comparative analysis of statistical methods to leverage external data and analyze randomized controlled trials. Multiple operating characteristics are discussed, such as the control of false positive results, power, and the bias of the treatment effect estimates, across candidate statistical methods. We compare the statistical methods through a broad  set of simulation scenarios. We then compare the methods using a collection of datasets with individual patient-level information from several glioblastoma studies in order to provide recommendations for future glioblastoma trials. 
\end{abstract}

\begin{keyword}
\kwd{External data}
\kwd{Clinical Trials}
\end{keyword}

\end{frontmatter}

\section{Introduction}

In recent years there has been significant interest in the integration of external information into clinical trial analyses, due to the growing availability of datasets as well as potential efficiency gains and cost reductions \citep{corrigan-curay_real-world_2018,khozin_real-world_2017}. Various methods have been proposed to augment the  analysis of clinical trials with external  data from previous clinical studies, disease registries, or electronic health records \citep{di_maio_real-world_2020, ghadessi_roadmap_2020}. External patient-level data provide useful information on the  outcome distributions under different therapies and pre-treatment patient profiles.
           
External data can  be used for multiple purposes, including (i) improving inference on treatment effects, (ii) reducing sample size, duration, and costs through better interim decisions during the trial, and (iii) enabling unbalanced randomization (e.g. 2:1 or 3:1) to increase the allocation to the experimental therapy, which can accelerate accrual \citep{dias_patient_2016} and  improve the characterization of the  toxicities under the experimental treatment  \citep{polley_leveraging_2024}. More broadly, external data can inform various aspects of a clinical trial: for example,  design decisions about sample size and eligibility \citep{thall_incorporating_1990}, interim analyses for sample size re-estimation or early stopping \citep{ventz_use_2022}, and final subgroup analyses at trial completion \citep{schwartz_harmonized_2023}. In addition, external data can be used throughout the clinical development, from phase 1 trials and early efficacy studies to  registration  trials \citep{yap_application_2022}. External data are particularly useful in trials where resources are limited, such as in rare diseases or when treatments are particularly costly \citep{mishra-kalyani_external_2022}.

External control data have been useful in a growing number of clinical trials. For detailed examples, \cite{jahanshahi_use_2021} survey non-oncology trials that used external data for FDA drug registration, \cite{mishra-kalyani_external_2022} reviews recent examples in oncology, and \cite{schwartz_clinical_2024} survey externally controlled trials reported on ClinicalTrials.gov. Most of the clinical studies discussed in these  reviews are  non-randomized, single-arm studies without a control group.
    
Here we compare candidate procedures to leverage external patient-level datasets (i.e. pre-treatment covariates and outcomes) in the analysis of randomized controlled trials (RCTs) with the goal of estimating and testing treatment effects of new therapeutics. Following the literature \citep{higgins_borrowing_1996, viele_use_2014, fu_covariate_2023}, we will use the term \textit{borrowing method} to refer broadly to any statistical method that allows external data to influence inference on the treatment effects. We focus on RCTs and on the use of external control data from multiple studies representative of the standard of care, such as the control data from clinical trials in a specific disease during the last 5 years. We compare several methods, including propensity score weighting \cite{freedman_weighting_2008} and random effects modeling \cite{allison_fixed_2009, dersimonian_meta-analysis_1986}, to estimate and test the treatment effects of the experimental therapeutics.

The use of external information in the analysis of RCTs presents several risks. Differences in the distribution of pre-treatment covariates across clinical studies and/or model misspecification, if not properly accounted for, lead to bias and inadequate control of false positive and negative results \cite{ventz_design_2022}. Moreover, unmeasured confounding (i.e. imbalance between populations of unmeasured pre-treatment characteristics that predict the outcome), is a potential component of heterogeneity and a source of bias that is difficult to account for. Additional challenges include potential measurement errors \cite{kobak_why_2007, morgan_effect_1987} and subtle differences in the definition of the outcomes across studies \cite{alkhaffaf_reporting_2018, broich_outcome_2007}. These and other characteristics of the available datasets determine the degree of difficulty and feasibility associated with the integration of external data in the analysis of RCTs. Indeed, in the absence of these concerns, external and RCT data could simply be pooled together. Crucially, the RCT's randomized internal control data offer the possibility of detecting potential distortion mechanisms --- such as unmeasured confounding --- in the external control data \citep{polley_leveraging_2024, fu_covariate_2023, viele_use_2014}.

There are several methods to account for the potential imbalance and distortion mechanisms that we mentioned. For example, propensity score procedures and semi-parametric regression models that link pretreatment covariates and outcomes can be effective solutions to account for different distributions of pre-treatment covariates across datasets; see for instance \cite{elze_comparison_2017} and \cite{chen_power_2000}. Propensity scores are typically used via weighting \citep{austin_moving_2015} and matching \citep{stuart_matching_2008, stuart_matching_2010, lim_minimizing_2018}, or included as a predictor in the outcome regression model \citep{hade_bias_2014}. Random effects models have been used to describe confounding mechanisms and, more generally, differences across studies \cite{dersimonian_meta-analysis_1986, graubard_regression_1994}. Other useful methods that combine ideas from the literature on propensity scores and random effects have been developed \cite{hong_evaluating_2006, arpino_propensity_2016, thoemmes_use_2011}; see also \cite{li_propensity_2013} and \cite{chang_propensity_2022}. In addition, several popular approaches to combine data from a clinical trial and external datasets are based on Bayesian modeling; we mention power priors \citep{chen_power_2000, ibrahim_power_2015, alt_scale_2023}, meta-analytic priors \citep{neuenschwander_summarizing_2010, schmidli_robust_2014}, commensurate priors \citep{hobbs_commensurate_2012, hobbs_adaptive_2013, murray_combining_2015}, multi-source   exchangeability priors \citep{kaizer_bayesian_2018, alt_leap_2023}, and elastic priors \citep{jiang_elastic_2023}. These methods formalize through prior distributions the available information from the external data, as well as the investigator's expectations, about potential distortion mechanisms that may cause artifacts and  systematic discrepancies between the trial control data and the external control  data.

We mention two motivations for our comparative analyses. First, sharing control data across concurrent RCTs can produce considerable advantages. For example, \cite{kotecha_prospectively_2022} discussed the concept of data-sharing networks of concurrent RCTs and produced encouraging results on efficiencies and risks of bias. Each RCT in the network is analyzed using both the data generated by the trial and shared control data from the other studies. The second motivation is that a  portion of early-stage cancer trials use  2:1 or 3:1 randomization. Often this choice is motivated by the goal of accelerating patients' enrollment \cite{featherstone_why_2002} and gathering information on the experimental therapy. The use of external control data in these trials is particularly attractive, as it can increase the likelihood of detecting treatment effects and improve the accuracy of the treatment effect estimates. These potential advantages are particularly relevant in precision oncology, where the sample sizes of biomarker-defined subgroups are often small \cite{eichler_randomized_2021}. 

We compare several methods for integrating external control data into the analysis of RCTs, using both simulations and a collection of glioblastoma (GBM) datasets. We describe the relative efficiencies and risks associated with each method using interpretable metrics, such as power, the probability of false positive results, bias, and the mean-squared error of the treatment effect estimates. In our simulations, we include ideal scenarios in which it is useful and straightforward to leverage information from external datasets, with minimal risks of bias. We then discuss a variety of scenarios with various distortion mechanisms, such as unmeasured confounders, in which the use of external data can translate into biased findings. In our simulation study, these distortion mechanisms impact the ranking of the candidate methods, from best to worst performances captured by standard metrics (e.g. bias). In other words, we provide a catalog of scenarios with potential distortion mechanisms and illustrate how the performances and rankings of candidate statistical plans --- including statistical plans that do not leverage external data --- vary across them. 

We also illustrate how data collections of completed studies can support the choice of an appropriate method for the analysis of future RCTs. In particular, we use a collection of individual patient-level data (IPLD) from completed GBM randomized studies that compared experimental treatments to the current standard of care (temozolomide + radiation) \cite{rahman_accessible_2023}. The IPLD include both outcome data and pre-treatment patient characteristics. A disease-specific data collection is useful to describe the relative merits and risks  of candidate statistical methods, in order to provide context-specific recommendations for future studies; see for example \cite{avalos-pacheco_validation_2023} and \cite{rahman_leveraging_2021}. Previous contributions assessed statistical methods and trial designs using a “leave-one-study-out” schema. We use this schema, and  in our implementation we subsample the control arm of a GBM trial, to mimic via simulations a GBM RCT that evaluates an experimental treatment with null effects. The  resulting {\it in silico} experimental and control arms are generated in the same way, by subsampling the control arm of a GBM trial. Then, the analysis of the {\it in silico} trial is augmented by other control arms in the GBM data collection from other studies (i.e. our external data). This exercise, repeated using various methods for data analysis and subsampling different control arms, allows us to provide context-specific comparisons of candidate methodologies. To emulate a GBM trial that tests an experimental drug with positive effects, we subsample the control arm of a completed GBM study from our data collection, and directly spike in treatment effects in the {\it in silico} experimental arm, for example by extending the survival times. In other words, we create simulation scenarios that include RCT and external data that are tailored to this disease setting by using subsampling procedures applied to a  collection of GBM datasets.

\section{Methods for analyzing RCTs and integrating external data}\label{sec:Methods}
We consider the analysis of an RCT, augmenting the data generated from the trial with multiple external datasets.
Patients in the external datasets and in the the RCT have the same  clinical 
condition.
Here we describe candidate methods (Subsections \ref{sec:TTP} to \ref{sec:psx}).
We will then compare the methods in Section \ref{sec:Comparisons}. 
The procedures that we evaluate provide inference on the average treatment effect $\tau$, 
which is   the mean difference 
between the outcomes under the treatment and control therapies in the trial population. 

Patients in the RCT have been $r:1$ randomized  to the experimental and control arms.
In what follows the control therapy will be the standard of care (SOC).
We use the index $i=1$ for the RCT that we analyze and $i=2,\ldots, I$ for the other studies.
The dataset generated from the RCT will be indicated as $D_1 
=\left(\bm X_{1j}, T_{1j}, Y_{1j} \right)_{i\le n_i}$. 
Patients are indexed by $j=1,\ldots,n_i$. 
The notation  $(\bm X,T, Y)$ indicates  pre-treatment patient characteristics (a vector of 
$q$ potential confounders), 
treatment indicators ($T_{ij}=0$ for the SOC, and $T_{ij}=1$ for the experimental treatment), and the outcomes. We will consider a binary outcome, treatment response $Y=0,1$. We define the treatment effect as
\begin{equation} \label{tx_eff_def}
    \begin{split}
            \tau &= E \left( Y_{i1} | T_{i1} = 1 \right) - E \left( Y_{i1} | T_{i1} = 0 \right),
    \end{split}
\end{equation}
This is the difference between the expected outcomes of treated and control patients, marginalizing with respect to the distribution of the covariates in the RCT population.
The external datasets $D_2,\ldots,D_I$
include IPLD and have the same structure as $D_1$, with the same pre-treatment covariates and outcomes. The external datasets will include information only on the control therapy (i.e., $T_{ij}=1$ when $i\ge2$).

In our comparisons, we will also consider RCT designs and statistical plans that \textit{do not} leverage external data.
We include two approaches: a Z-test for proportions (ZPROP) to compare  experimental and control therapies and a generalized linear models (GLM). They ignore the external datasets $D_2\ldots D_I$. The Z-test for proportions  ignores also the covariates $\bm X$, while the GLM, which  in our comparisons is a  logistic regression analysis, includes them. The list of methods that we discuss is not exhaustive, but it includes popular approaches and covers a range of different strategies for integrating external data into the analysis of an RCT, including weighting, regression, hierarchical modeling, and stratification.

\subsection{Test-then-pool (TTP)}\label{sec:TTP}

We summarize the two-step \textit{test-then-pool} (TTP) procedure discussed in \cite{viele_use_2014}. 

{\it Step 1: Select datasets.}
The first step provides a list of datasets that will be included the analysis of the RCT. 
In other words, the output  is a subset $\mathcal{S}$ of $\{1,\ldots, I\}$, 
which includes the RCT ($i=1$) that we analyze, but  might exclude some of the external studies $D_2,\ldots D_I$
if
the study-specific  ($i=2,\ldots,I$) outcome distribution appears incongruent with the outcome data from the control arm of the RCT  $D_1$.
In particular study $i=2,\ldots,I$ is included or excluded based on hypothesis testing. 
The procedure tests the null hypothesis 
of identical response probabilities under the SOC in the $i$-th dataset ($i=2,\ldots,I$) and in the RCT population. In our comparisons we use a standard two-sample Z-test for proportions, ignoring the pre-treatment characteristics $\bm X$, with a significance level of 0.2 (a candidate threshold considered in \cite{viele_use_2014}). Since the TTP procedure does not adjust for pre-treatment confounders $\bm{X}$, one expects biased estimates when the marginal distributions of $\bm{X}$ in the RCT controls and the external data are different (see for example Scenario 5 in Table 1 and Figure 2 below).

{\it Step 2: Pool and analyze.} The selected  datasets   $(i\in \mathcal{S})$ are  combined into a single data matrix $\tilde{D}$.
Inference on the treatment effect $\tau$ is based on a standard analysis of $\tilde{D}$, 
which does not make distinctions between outcomes and patients treated with the SOC from 
the RCT and the external studies.
This  second step  ignores  the selection mechanism of \textit{Step 1}. 
The  estimate $\hat \tau$ is the mean difference in $\tilde{D}$ of the outcomes under the experimental and control therapies:
$$\hat \tau=\frac{\sum_{i\in\mathcal{S}}\sum_j Y_{ij} \times T_{ij}}{\sum_{i\in\mathcal{S}}\sum_j T_{ij}}-
\frac{\sum_{i\in\mathcal{S}}\sum_j Y_{ij} \times (1-T_{ij})}{\sum_{i\in\mathcal{S}}\sum_j (1-T_{ij})}.
$$
Similarly, 
the null hypothesis $(H_0: \tau=0)$ is tested using a standard  $Z$-statistic  
to  summarize the  available  evidence 
from $\tilde{D}$
of  different outcome distributions under the experimental and control therapies. 

\subsection{Propensity score weighting (PSW)}\label{sec:PSW}

In this section we describe a propensity score procedure to analyze an RCT, augmenting $D_1$ with external data $D_2,\ldots, D_I$. The procedure involves three steps: 1. calculating propensity scores, 2. maximizing a weighted log-likelihood function, and 3. treatment effect estimation and hypothesis testing. 

\medskip

{\it Step 1: Calculate propensity scores $e_{ij}$.}
We combine the patient characteristics $\bm X_{ij}$ from all datasets $D_1,\ldots,D_I$ into a single data matrix 
$\bar{D}$ of dimension $(\sum_i n_i) \times (q+1)$, where $q$ is the number of pre-treatment characteristics. 
Each row of $\bar{D}$ is dedicated to one patient in the data collection $D_1,\ldots,D_I$. 
The first $q$ columns provide the   characteristics $\bm X_{ij}$, while the last entry
is equal to 1 if the patient participated in the RCT and 0 otherwise. 
The scores $e_{ij}=e(\bm X_{ij})$ are obtained through a binary regression with the first $q$ columns of $\bar{D}$ 
used as covariates and the last column as  dependent variable. In our implementation 
we use logistic regression and maximum likelihood estimation of the coefficients. 
The result is a regression function $e: \mathbb{R}^q \rightarrow (0,1)$. 
\medskip

{\it Step 2: Maximize the weighted log-likelihood.}
We model the relation between outcomes $Y_{ij}$, covariates $X_{ij}$, and treatment $T_{ij}$
using a generalized linear model with parameters $\beta$ and $\gamma$:
\begin{align}
	P( Y_{ij}=1 \mid \bm X_{ij}, T_{ij})= F\left(\beta_0+\bm X_{ij}^T\beta + \gamma T_{ij} \right),\\\label{eq:pswglm}
	 \text{ for } i=1 \ldots I, \text{ and }j=1 \ldots n_i.\nonumber
\end{align} 
Throughout the manuscript we use $F$ to denote the logistic function, $F(t)=\left(1+e^{-t}\right)^{-1}$.

Then, we can use  the  propensity scores for weighting each term in the log-likelihood,
\begin{align}
	&\ell(\beta_0,\beta,\gamma ; D_1\ldots D_I)=\sum_{i=1}^I \sum_{j=1}^{n_i}\left[
	Y_{ij} \left(\beta_0+\boldsymbol{X}_{ij}^T \beta+\gamma T_{ij}\right)\right.\\ \nonumber 
	&\left.-\log \left(1+\exp{\left(\beta_0+\boldsymbol{X}_{ij}^T \beta+\gamma T_{ij}\right)}\right) \right]\times \left(C {\frac{e_{ij}}{1-e_{ij}}}\right)^{\mathbb{I}\left[i \neq 1\right]},\\ \label{eq:psw}
	&\text{ for }i=1\ldots I, j=1\ldots n_i, \nonumber
\end{align}
where $\mathbb{I}$ is an indicator function that  assigns  weights equal to 1 to the individual records from the trial and  weights proportional to $\frac{e_{ij}}{1-e_{ij}}$ to the others. 
We estimate $\beta_0$, $\beta$ and $\gamma$ by maximizing $\ell(\beta_0,\beta,\gamma ; D_1\ldots D_I)$.
The constant $C$ modulates the relative influence 
of the RCT data and the external data on on  the  estimation of $\tau$. 
The role of $C$ is similar to the power parameter used in the power-prior framework,
which can be used    to analyze RCTs  integrating data from previous studies   \cite{ibrahim_power_2000}.
In our comparative study we will set $C = 1$.  Figure S1 in the Supplementary Material illustrates results for $C = 0.5 $ and $ 0.1$ \cite{kotecha_supplement_sims}.

\medskip

{\it Step 3: Estimate treatment effect and hypothesis testing.}
The estimates $(\hat\beta_0, \hat\beta, \hat\gamma$) can then be used to estimate $\tau$
$$\hat\tau= n_1^{-1} \sum_j \left [F\left(\hat\beta_0+ \bm X_{1j}^T\hat\beta +\hat\gamma\right) - 
F\left(\hat\beta_0+\bm X_{1j}^T\hat\beta \right) \right ].$$ 
\sloppy To estimate the variance of 
$\hat\tau$ or
$\hat\theta=(\hat\beta_0,\hat\beta,\hat\gamma)$,   for confidence intervals or hypothesis testing, we can use the sandwich estimator \cite{robins_marginal_2000, austin_moving_2015}. 
Consider $\ell(\beta_0,\beta,\gamma ; D_1\ldots D_I)$,  let $\nabla^2 \ell\left(\theta ; D_1, \ldots D_I\right)$ denote the matrix of second-order derivatives, and $g_{ij}(\theta)$  the   partial derivatives \begin{align}
	&\nabla \left[
Y_{ij} \left(\beta_0+\boldsymbol{X}_{ij}^T \beta+\gamma T_{ij}\right)\vphantom{\left(C {\frac{e_{ij}}{1-e_{ij}}}\right)^{\mathbb{I}\left[K_{ij} \neq 1\right]}}\right.\\\nonumber
&\left.-\log \left(1+\exp{\left(\beta_0+\boldsymbol{X}_{ij}^T \beta+\gamma T_{ij}\right)}\right) \left(C {\frac{e_{ij}}{1-e_{ij}}}\right)^{\mathbb{I}\left[K_{ij} \neq 1\right]}\right]\end{align} of 
the weighted log-likelihood components.
The sandwich estimator \cite{freedman_so-called_2006} of the variance of $\hat{\theta}$ is $\hat V=(-A)^{-1}B(-A)^{-1}$, where
$$A=\nabla^2\left.\ell\left(\theta ; D_1\ldots D_I\right)\right\rvert_{\theta=\widehat{\theta}}$$ and 
$$
\begin{aligned}
	B&=\sum_{i=1}^I \sum_{j=1}^{n_i} \left[ g_{ij}(\widehat\theta)\right]\left[ g_{ij}(\widehat\theta)\right]^T.\\
\end{aligned}
$$
We use $\widehat{\sigma}_{PSW, \gamma}^2$ to indicate the estimated variance of $\hat\gamma$. The bootstrap is  an alternative methodology that can be used to estimate the variance of $\hat\theta$. In both cases the null hypothesis ($H_0:\hat\gamma\leq 0$) can be tested using the statistic $\frac{\hat\gamma}{\sqrt{\widehat{\sigma}^2_{PSW, \gamma}}}$. We will focus primarily on inference for $\gamma$ (the odds ratio), though inference on $\tau$ (the difference of means) can be carried out similarly through standard approaches such as  bootstrapping.


\subsection{Fixed effects model (FE)}\label{sec:FE}
We consider a fixed effects model:
\begin{align} \label{eq:FE}
	P(Y_{ij}=1|\delta, \beta, \gamma, \boldsymbol{X}_{ij}, T_{ij}) = F(\delta_i+\boldsymbol{X}_{ij}^T \beta +\gamma T_{ij}), \\\nonumber
	\text{ for } i=1 \ldots I, \; j=1\ldots n_i,
\end{align}
where $\delta=(\delta_1,\ldots,\delta_I)$. 
The model includes study-specific intercepts  $\delta$, 
while the regression coefficients $\beta$ that link individual covariates $\bm X_{ij}$ 
to the outcome distribution are identical across studies. 
Fixed effects models are often used for regression analyses with clusters of subjects
\cite{mcculloch_generalized_2001, wooldridge_econometric_2010, hsiao_analysis_2022}.

The estimate of the treatment effect $\tau$ is 
\begin{equation} \label{fe_tau_hat}
    \hat\tau = n_1^{-1} \sum_j\left[F \left(\hat{\delta}_1+\boldsymbol{X}_{1 j}^T \hat{\beta}+\hat{\gamma}\right)-F\left(\hat{\delta}_1+ \boldsymbol{X}_{1 j}^T \hat{\beta}\right)\right],
\end{equation}
where $\hat\delta_1$, $\hat \beta$ and $\hat\gamma$ are maximum likelihood estimates.

%
The Hessian of the log-likelihood 
function at $(\hat\delta_1, \hat \beta, \hat\gamma)$
allows us to 
directly estimate  the variance-covariance matrix of  the  parameters $(\hat\delta,\hat{\beta},\hat{\gamma})$.
We use the  estimated variance $\widehat {\sigma}^2_{FE,\gamma}$ of $ \hat \gamma$ 
%
to test the null hypothesis $\left(H_0: \gamma \leq 0 \right)$, using the test statistic $\hat \gamma / \widehat {\sigma}_{FE,\gamma}$.

\subsection{Random effects model (RE)}\label{sec:RE}
We discuss the use of the random effects model
\begin{align}\label{eq:RE}
	P(Y_{ij}=1|\beta_0,\delta, \beta, \gamma, \boldsymbol{X}_{ij}, T_{ij})&= F(\beta_0+\delta_i+\boldsymbol{X}_{ij}^T \beta +\gamma T_{ij}),\\\nonumber
	&\text{ for }i=1 \ldots I ,\; j=1\ldots n_i,\\ \nonumber
	\delta_i &\sim N\left(0, \sigma_\delta^2\right),\nonumber
\end{align}
where $\beta_0$ is the common intercept, and $\delta=(\delta_1\ldots \delta_I)$ are study-specific random effects.
To estimate the parameters $\theta=(\beta_0, \beta, \gamma)$ and $\sigma_\delta^2$, 
we focus on the marginal likelihood, integrating out the random effects. 

\medskip
We estimate the parameters $\theta$ and $\sigma_\delta^2$ of the random effects model (\ref{eq:RE}) using the penalized likelihood 
\begin{align}\label{eq:2}
	&L(\sigma_\delta, \theta; D_1,\ldots, D_I)\\\nonumber
	&=
	g(\sigma_\delta)
	\prod^I_{i=1}
	\int_{-\infty}^{\infty}
	\prod_{j=1}^{n_i}
	P(Y_{ij}| \delta_i, \theta,\boldsymbol{X}_{ij}, T_{ij}) \phi (\delta_i/ \sigma_\delta) /\sigma_\delta
	\; d\delta_i. 
\end{align}
where $\phi$ indicates the standard normal density. 
In particular we use a gamma distribution $g(\sigma_\delta) \propto \sigma_\delta^{\eta-1} e^{-\lambda \sigma_\delta}$, with $\eta = 2$ and $\lambda = 0.01$ following the recommendation of \cite{chung_nondegenerate_2013}. This choice makes $g(\sigma_\delta) \approx \sigma_\delta$ and ensures that the penalized MLE of $\sigma^2_\delta$ is strictly positive. We estimate the parameter values in (\ref{eq:2}) via penalized maximum likelihood: 
\begin{equation}\label{eq:RE_ests}
	(\hat\sigma_\delta, \hat \theta)= \underset{\sigma_\delta, \theta}{\arg\max}\; L(\sigma_\delta, \theta; D_1,\ldots, D_I).
\end{equation} 
See \cite{demidenko_mixed_2013, chung_nondegenerate_2013} for an analysis of statistical properties of these estimates. We compute $(\hat \sigma_\delta, \hat\theta)$ and then estimate $\delta_1$ and $\tau$:
$$\hat \delta_1 =\underset{
	\delta_1}{\arg\max} \left[
\phi\left(\delta_1 / \hat\sigma_\delta\right) / \hat\sigma_\delta
\times \prod_{j=1}^{n_1}
P(Y_{1j}| \delta_1, \hat \theta, \boldsymbol{X}_{1j}, T_{1j})\right], $$
\begin{equation} \label{re_tau_hat}
    \begin{aligned}
        \hat\tau &= n_1^{-1} \sum_j \Big[ F \left(\hat\beta_0+\hat{\delta}_1+\boldsymbol{X}_{1 j}^T \hat{\beta}+\hat{\gamma}\right) \\
        &\hspace{5.5em} - F\left( \hat\beta_0+\hat{\delta}_1+\boldsymbol{X}_{1 j}^T \hat{\beta}\right) \Big].
    \end{aligned}    
\end{equation}

The estimate
$\hat\theta$ is   approximately normally distributed when $I$ and $\min_i n_i$ diverge \cite[Chapter~3.6.2]{demidenko_mixed_2013}. 
To estimate the variance of  $\hat\gamma$ we  compute the  second derivative of the penalized marginal log-likelihood $\log L(\sigma_\delta, \theta; D_1,\ldots, D_I)$ at $(\sigma_\delta, \theta)=(\hat\sigma_\delta, \hat \theta)$. 
We then use  the resulting estimate of the variance $\widehat{\sigma}_{RE,\gamma}^2$   and  the standardized statistic $\hat\gamma/\widehat{\sigma}_{RE,\gamma}^2$, which is approximately normally distributed, to test the 
null hypothesis $H_0: \gamma \leq 0 $.

\subsection{Propensity score stratification and random effects (PSS-RE)}\label{sec:psx}
This section describes a procedure that combines propensity score stratification and random effects (PSS-RE). 
The procedure has two main components that we indicate as steps.

\medskip

{\it Step 1. Augment the RCT data.} 
The approach leverages an 
augmented dataset $\tilde {D}$, which consists of the RCT data $D_1$ and subsets $D_{i,sub}$ of the external datasets $D_{i}, i = 2\ldots I$. The subsets $D_{2,sub}, \ldots, D_{I,sub}$ are selected 
to have approximately the same distribution of the propensity scores as $D_1$. We repeat the following three operations for
each dataset $D_{i}, i = 2\ldots I$, one at a time. 
\medskip

{\it (a) Calculate propensity scores.} We combine the RCT data $D_1$ and the external data of study $i$ into a single dataset $\widetilde D_{i}$. 
This is a $(n_1+n_{i})$ by $(q+1)$ data-matrix, where the first $n_1$ rows correspond to IPLD in the RCT and the remaining $n_i$ rows correspond to the IPLD from the i-th external dataset. 
The first $q$ entries of each row provide patient pre-treatment characteristics $\bm X_{ij}$, while the last entry equals 1 if the patient was enrolled into the RCT and 0 otherwise. We then fit a binary regression model ($e^{i}$) with the first $q$ columns of $\widetilde D_{i}$ used as covariates and the last column as 
dependent variable. We use logistic regression and maximum likelihood estimation of the coefficients. This allows us to 
compute the propensity scores $e^{i}_{\ell, m} = e^{i}(\bm X_{\ell m}),$ for $m =1, \ldots, n_\ell$ and $\ell=1,i$.
\medskip

{\it (b). Stratify.} We then stratify patients in $D_1$ into $K$ subgroups of approximately equal size
defined by the $(k/K)$-quantiles (indicated by $Q_{i,k}$, $k=0,\ldots, K) $ of the 
scores $\{e^{i}_{1,m}\}_{m=1}^{n_1}$. We use $K = 5$ in our simulations, as this parametrization is often adopted in the literature and was proposed and discussed by \cite{cochran_effectiveness_1968},  \cite{rosenbaum_central_1983}, and \cite{rosenbaum_reducing_1984}. These studies observed that as $K$ increases, treatment effect estimates tend to exhibit higher variance and lower bias. As discussed in Section 4, analysts can assess the impact of $K$ on the OCs  --- focusing on a specific trial ---  through tailored simulations. \medskip

{\it (c). Select patients in $D_i$.} 
Let $L_i\geq 0$ be the largest integer such that we can count  $L_i$ or more patients in dataset $D_i$ with propensity scores 
$e^{i}_{\ell} \in ( Q_{i,k-1}, Q_{i,k}]$ for every stratum $k=1, \ldots, K$.
If $L_i>0$ then we augment the RCT data with $K \times L_i$ individual patient-level data-points from $D_i$, i.e. including $L_i$ patients from $D_i$ with propensity scores in $( Q_{i,k-1}, Q_{i,k}]$ for each stratum $k=1, \ldots, K$.
\medskip

{\it Step 2. Treatment effect estimation and hypothesis testing.} Once {\it Step 1} has been applied to each external dataset $D_{i},i\in 2\ldots I$, we apply the random effects model (Equation \ref{eq:RE}) and the procedure described in subsection \ref{sec:RE} to analyze the
data matrix $\tilde {D}$, which includes $D_1$, the  subsets $D_{2,sub}\ldots D_{I,sub}$ (the output of \textit{Step 1}), and the outcomes of the patients that were selected. 
Hypothesis testing and treatment effect estimation are performed as described in \ref{sec:RE}, simply replacing $D_1\ldots D_I$ with $ \tilde D$. 

\subsection{Propensity scores within random effects models (PS-RE)}\label{sec:PSC}

We consider another method that combines propensity scores and random effects modeling. 
We compute propensity scores \cite{imbens_role_2000} and include them as covariates in the outcome regression model. 
The procedure contains two main steps.

{\it Step 1. Calculate generalized propensity scores $ e^G_{ij}$}. 
The generalized propensity score  $e_{i j}^G$ is an extension of propensity scores to multiple groups or datasets. To calculate $e_{i j}^G$, we define the $\sum_{i=1}^I n_i$ by $q+1$ matrix 
$$
\bar {D}=
\begin{bmatrix*}[l] 
	\bm X_1, \bm 1\\
	\bm X_2, \bm 2\\
	\ldots, \ldots\\
	\bm X_I, \bm I\\
\end{bmatrix*} 
$$
Each row of $\bar {D}$
corresponds to IPLD for one individual in the data collection. 
The first $q$ entries of each row provide the individual covariate profile $\bm X_{ij}$, while the last entry of each row identifies   the dataset of origin $i\in \{1\ldots I\}$ for each patient. 
We estimate the generalized propensity scores  based on a multinomial logistic regression, using the first $q$ columns of $\bar {D}$ 
as covariates and the final column as dependent variable. 
We estimate the regression parameters through maximum likelihood. 
The result is a regression function $e^G$ that maps $R^q$ 
into the  $(I-1)$-dimensional simplex. 
The individual propensity score vector $e^G_{i j}$ is equal to $e^G\left(\boldsymbol{X}_{i j}\right)$.
\medskip

{\it Step 2. Include $e^G_{ij}$ in a random effects regression model.} 
We include the covariates $\bm X_{ij}$ and random effects terms $\delta_i$ in the outcome regression model:
\begin{align}\label{eq:ra}
	&P(Y_{ij}=1\mid\beta_0,\delta,\beta, \gamma, \boldsymbol{X}_{ij}, T_{ij},\zeta,e^{G}_{ij} )\\\nonumber&=F\left(\beta_0+\delta_i+\boldsymbol{X}_{ij}^T \beta +\gamma T_{ij}+G(e^{G}_{ij}, \zeta) \right),\\
	\delta_i&\sim N(0,\sigma^2_\delta),\nonumber
\end{align}
where $\delta=\left(\delta_1 \ldots \delta_I\right)$ and $G(\cdot)$ is
a  function which captures the association 
between propensity scores and
outcomes  parameterized by $\zeta$. 
The inclusion in regression models without random effects  of propensity scores, potentially combined with other prognostic variables, has been discussed  and substantial limitations have been identified in the literature \cite{garrido_covariate_2016, hade_bias_2014}.  One important concern is that misspecification of the regression model can result in bias. Although in our setting  we have randomization in $D_1$ and study-specific random effects $\delta$, the same concern remains, and in particular is nearly identical when the external datasets $D_1,\ldots,D_I$ present similar distributions and support the hypothesis that $\sigma^2_\delta \approx 0 $.   Despite this and other weaknesses, multiple systematic reviews showed that the  approach, i.e. the   inclusion  in regression models of propensity scores, is frequently used  in the medical literature \cite{weitzen_principles_2004, shah_propensity_2005, sturmer_review_2006}. In our implementation $G(e^G_{ij}, \zeta)= \sum_k \zeta^{(k)} \log \frac{e_{ij}^{G(k)}}{1-e_{ij}^{G(k)}}$, where $\zeta^{(k)}$ and $e_{ij}^{G(k)}$ are the components of $\zeta \in \mathbb R^I$ and $e_{ij}^G$ respectively. The estimation of the model parameters, inference on $\gamma$, treatment effect estimation and hypothesis testing are carried out as described in Section \ref{sec:RE}.

\begin{landscape}
	%
	\begin{table}[htbp!]
		\resizebox{24cm}{3.2cm}
		{%
			\Large
			\renewcommand{\arraystretch}{1.06}
			\begin{tabular}{p{10.5cm}llllllllllll}
				\hline
				\diagbox[width=10.5cm,height=1.3\line]{\textbf{Feature}}{\textbf{Scenario}} & \textbf{1} & \textbf{2} & \textbf{3} & \textbf{4} & \textbf{5} & \textbf{6} & \textbf{7} & \textbf{8} & \textbf{9} & \textbf{10} & \textbf{11} & \textbf{12} \\\hline \hline
				{\textbf{A.} Sample size of the RCT, $n_1$} & 100 & \cellcolor{yellow}120 & \cellcolor{yellow}80 & 100 & 100 & 100 & 100 & 100 & 100 & 100 & 100 & 100 \\\hline
				{\textbf{B.} Sample size of the external datasets} & 25 & \cellcolor{yellow}30 & \cellcolor{yellow}20 & 25 & 25 & 25 & 25 & 25 & 25 & 25 & 25 & 25 \\\hline
				{\textbf{C.} Number of datasets $I$} & 4 & \cellcolor{yellow}2 & \cellcolor{yellow}8 & 4 & 4 & 4 & 4 & 4 & 4 & 4 & 4 & 4 \\\hline
				{\textbf{D.} RCT randomization ratio $r:1$} & 2:1 & 2:1 & 2:1 & 1:1\cellcolor{yellow} & 2:1 & 2:1 & 2:1 & 2:1 & 2:1 & 2:1 & 2:1 & 2:1 \\\hline
				{\textbf{E.} Covariate Distributions $f_i$ across datasets} & Identical & Identical & Identical & Identical & Heterogeneous \cellcolor{yellow} & Identical & Heterogeneous \cellcolor{yellow} & Identical & Heterogeneous \cellcolor{yellow} & Identical & Identical & Identical \\\hline
				{\textbf{F.} Value of the intercepts $\delta_i$ across datasets} & Identical & Identical & Identical & Identical & Identical & Heterogeneous\cellcolor{yellow} & Heterogeneous\cellcolor{yellow} & Identical & Identical & Identical & Heterogeneous \cellcolor{yellow} & N/A \\\hline
				{\textbf{G.} The unmeasured covariate affects 
					the response probabilities} & No & No & No & No & No & No & No & Yes\cellcolor{yellow} & Yes\cellcolor{yellow} & No & No & No \\\hline
				{\textbf{H.} Parameter vectors $\beta_i$ across datasets} & Identical & Identical & Identical & Identical & Identical & Identical & Identical & Identical & Identical & Heterogeneous\cellcolor{yellow} & Identical & N/A \\\hline
				{\textbf{I.} Model Specification} & Correct & Correct & Correct & Correct & Correct & Correct & Correct & Correct & Correct & Correct & Correct & 
				Incorrect \cellcolor{yellow}\\
				\hline\multicolumn{13}{l}{\parbox[l][1cm][c]{18cm}{ }}\\[-2ex]	\hline
				{\textbf{J.} Marginal probability of $Y=1$: $D_1$, $T=0$ } & 0.39 (0.19) & 0.39 (0.19)& 0.39 (0.19)& 0.39 (0.19)& 0.39 (0.19)& 0.39 (0.19) & 0.39 (0.19
				) & 0.25 (0.16) & 0.25 (0.19) & 0.39 (0.21) & 0.65 (0.18) & 0.40 (0.15)
				\\\hline
				{\textbf{K.} Marginal probability of $Y=1$: $D_2$, $T=0$} &0.39 (0.19) & 0.39 (0.19) & 0.39 (0.19)& 0.39 (0.19)& 0.34 (0.18)& 0.28 (0.17)& 0.25 (0.19)& 0.25 (0.16) & 0.29 (0.16) & 0.39 (0.18)& 0.40 (0.15)
				& 0.40 (0.15)
				\\\hline
				{\textbf{L.} Marginal probability of $Y=1$: $D_1$, $T=1$} &0.66 (0.13)& 0.66 (0.13)& 0.66 (0.13) & 0.66 (0.13)& 0.66 (0.13) & 0.66 (0.13) & 0.66 (0.23) & 0.45 (0.14) & 0.45 (0.25) & 0.66 (0.16) & 0.85 (0.10)
				& 0.66 (0.15)
				\\\hline
				\multicolumn{13}{c}{}\\
		\end{tabular}}
		\resizebox{22cm}{!}{%
			\centering
			\small
			\begin{tabular}	{lllllllllllll}
				\hline
				\textbf{Feature} & \multicolumn{12}{c}{\textbf{Description}} \\ \hline \hline	
				
				{\textbf{A.} Sample size of the RCT, $n_1$ 
				} & 
				\multicolumn{12}{l}{\parbox[l]{18cm}{
						The number of patients enrolled in the RCT.}}\\ \hline
				{\textbf{B.} Sample size of the external datasets 
				} 
				& 
				\multicolumn{12}{l}{\parbox[l]{18cm}{ 
						The values of $n_2=n_3=\ldots =n_I$, the size of each external dataset. 
				}}\\ \hline
				{\textbf{C. } Number of datasets $I$ 
				} 
				& 
				\multicolumn{12}{l}{\parbox[l]{18cm}{ 
						The number of datasets, including the RCT dataset ($D_1$) and the external datasets ($D_2\ldots D_I$). }}\\ \hline
				{\textbf{D.} RCT randomization ratio $r:1$ 
				} & 
				\multicolumn{12}{l}{\parbox[l]{18cm}{ 
						The ratio of patients allocated to experimental and control arms. }}\\ \hline
				\parbox[l][][c]{7cm}
				{\textbf{E.}  Covariate Distributions $f_i$ across datasets 
				} 
				& 
				\multicolumn{12}{l}{\parbox[l][1.4cm][c]{18cm}{
						If identical, then $\bm P_{ij}=\left[P\left(X_{ij}^{(1)}=1\right), P\left(X_{ij}^{(2)}=1\right), P\left(X_{ij}^{(3)}=1\right)\right]=[0.4,0.5,0.5], i=1\ldots I,j=1\ldots n_i$ 
						(Scenarios 1-4, 6, 8, 10-11). 
						If Heterogeneous (Scenarios 5, 7 and 9), then $ \bm P_{ij}=[0.4,0.5,0.5],i=1,j=1\ldots n_i$; $\bm P_{ij}=[0.3,0.8,0.2],i=2,j=1\ldots n_i$;
						$\bm P_{ij}=[0.3,0.7,0.9],i=3,j=1\ldots n_i$;
						$\bm P_{ij}=[0.1,0.7,0.8],i=4,j=1\ldots n_i$. }} \\ \hline
				\parbox[l][][c]{6cm}
				{\textbf{F.} Value of the intercepts $\delta_i$ across datasets 
				} & 
				\multicolumn{12}{l}{\parbox[l][1.2cm][c]{18cm}{ 
						If identical, then $\delta_i=-0.4,i= 1\ldots I$ (Scenarios 1-5, 8-10). 
						If Heterogeneous then 
						$\displaystyle\delta=[\delta_1,\delta_2,\delta_3,\delta_4]=[-0.4,-0.9,-0.2,-0.6]$ (Scenarios 6 and 7) or $\displaystyle\delta=[0.7,0.5,0.5,-0.5]$ (Scenario 11). If N/A (Scenario 12) then  the conditional distribution $P(Y_{ij}=1\mid\bm X_{ij}=\bm x_{ij},T_{ij}=t_{ij})$ is identical across datasets $i=1\ldots I$.} }\\ \hline
				\textbf{G.} The unmeasured covariate affects 
				the response probabilities 
				& 
				\multicolumn{12}{l}{\parbox[l][1.35cm][c]{18cm}{ 
						If ``No'' (Scenarios 1-7, 10-12), then the outcome is independent of the unmeasured covariate $X^{(3)}$. In Scenarios 1-7 and 10-11, $\mu_\beta^{(3)}=\sigma_\beta^{(3)}=0$ in model \ref{eq:dgm}, where $\mu_\beta^{(3)}$ and $\sigma_\beta^{(3)}$ are the third components of $\mu_\beta$ and $\sigma_\beta$ respectively. If ``Yes'' (Scenarios 8-9) then $\mu_\beta^{(3)}=-1.8$ and the value of $\sigma_\beta$ is provided, Feature \textbf{H}. 
				} }\\ \hline
				\parbox[l][][c]{7cm}{
					\textbf{H.} Parameter vectors $\beta_i$ across datasets 
				} &
				\multicolumn{12}{l}{\parbox[l][2cm][c]{18cm}{ 
						If identical (Scenarios 1-9), then in model \ref{eq:dgm}, $\sigma_\beta=[0,0,0]'$ and $\beta_1=\beta_2=\ldots \beta_I=\mu_\beta$. Either $\mu_\beta=[0.5,-0.5,0]$ (Scenarios 1-7) or $\mu_\beta=[0.5,-0.5,1.8]$ (Scenarios 8-9). If ``N/A'' (Scenario 12) then $P(Y_{ij}=1\mid\bm X_{ij}=\bm x_{ij},T_{ij}=t_{ij})$ is identical across datasets $i=1\ldots I$. 
						If different, the $\beta_i$ vary across datasets. 
						They are randomly generated from the normal distribution described in model \ref{eq:dgm} with mean $\mu_\beta$ and the first two components of $\sigma_\beta$ equal to 0.8. The third component of $\sigma_\beta$ is 0 (if Feature \textbf{G}=``No'', Scenario 11) or 0.8 (if Feature \textbf{G}=``Yes''). 
				}} \\ \hline	
				\textbf{I.} Model Specification 
				& 
				\multicolumn{12}{l}{\parbox[l][0.85cm][c]{18cm}{ 
						If correct (Scenarios 1-11), then the outcome data are generated from model \ref{eq:dgm}, a logistic model without interaction terms between the covariates. If incorrect (Scenario 12), then the outcome data are generated according to the values in Table \ref{tab:scenario12}.}}\\ 			
				\hline\multicolumn{12}{l}{\parbox[l][0.6cm][c]{18cm}{ }}\\[-2ex]	\hline
				
				\textbf{J.} Marginal probability of $Y=1$: $D_1$, $T=0$ 
				& 
				\multicolumn{12}{l}{\parbox[l][0.8cm][c]{18cm}{We report the expectation and standard deviation of the proportion of positive responses in patients allocated to SOC in the RCT.}}\\ \hline
				\textbf{K.} Marginal probability of $Y=1$: $D_2$, $T=0$ 
				& 
				\multicolumn{12}{l}{\parbox[l][0.8cm][c]{18cm}{We report the expectation and standard deviation of the proportion of positive responses in patients allocated to SOC in $D_2$.}}\\ \hline
				\textbf{L.}  Marginal probability of $Y=1$: $D_1$, $T=1$ 
				& 
				\multicolumn{12}{l}{\parbox[l][0.8cm][c]{18cm}{We report the expectation and standard deviation of the proportion of positive responses in patients allocated to the experimental therapeutic in the RCT.}}\\ \hline
			\end{tabular}%
		}

		\caption{
			We generated RCT data using 12  scenarios (columns) with different features (rows \textbf{A.} to \textbf{I.}). 				
		}	\label{tab:scenarios}
	\end{table}
\end{landscape}

\section{Comparative analyses}\label{sec:Comparisons}

We examine the methods in Section \ref{sec:Methods} using simulations (subsection \ref{sec:sims}) and 
a collection of glioblastoma datasets (subsection \ref{Sec:Application}). 

\subsection{Simulations}\label{sec:sims}
We first define ideal simulation scenarios, without unmeasured confounding or other distortion mechanisms, in which it is effective to leverage information from external data without substantial risks of bias. Then we specify scenarios with various distortion mechanisms to evaluate the bias and other operating characteristics (OCs) when potential departures from standard assumptions occur, for instance model misspecification. Through these scenarios, we investigate the extent to which the OCs (e.g., accuracy of treatment effects estimates) deteriorate when we apply the candidate methods in section \ref{sec:Methods}. We evaluate if some of these methods are particularly robust to specific distortion mechanisms but less to others. In these scenarios we include different covariate distributions across datasets, study-specific intercepts of the outcome models, different covariate effects across datasets, and model misspecification.

\subsubsection{Simulation Model}
Recall, the RCT dataset $D_1$ consists of ($\bm X_{1j}$, $T_{1j}$, $Y_{1j}$), with patients indexed by $j=1\ldots n_1$. The external datasets $D_2, \ldots, D_I$ have the same structure.
In the 
simulations all patients in the external datasets receive the SOC (i.e., $T_{ij}=0$ when $i=2,\ldots, I$). 

In our scenarios the vector of pretreatment covariates $\bm X$ has three binary components, $X^{(1)}$, $X^{(2)}$ and $X^{(3)}$. Individual covariates $X_{ij}$ are generated independently from the other patients, potentially with different distributions $f_i$ across datasets. In each dataset the covariates $X^{(1)}$, $X^{(2)}$ and $X^{(3)}$ are independent. The first two covariates ($X^{(1)}$ and $X^{(2)}$) are observed by the investigators and affect the probability of a positive outcome ($Y=1$). The third covariate $X^{(3)}$ is not observed by the investigator, although it affects in several of our scenarios (Table \ref{tab:scenarios}) the probability of a positive outcome. The treatment assignments $T_{1j}$ in the RCT are randomly generated, with a ratio of $r:1$ for the experimental and control therapies. In all simulation scenarios (except Scenario 12), the outcomes $Y$ are generated from the following model, 
\begin{align}\label{eq:dgm}
	P\left(Y_{ij}=1 \mid \delta, \beta_i, \gamma, \boldsymbol{X}_{ij}, T_{ij}\right)&=F\left(\delta_i+\boldsymbol{X}_{ij}^T \beta_i+\gamma T_{ij}\right),\\
	\beta_i&\sim N(\mu_\beta,\text{diag}(\sigma_\beta)), \nonumber
\end{align}
where $\delta=(\delta_1\ldots\delta_I)$ and $F$ is the logistic function. The vectors $\beta_i$ can vary across datasets and simulations. The components $\beta_i^{(1)}$, $\beta_i^{(2)}$, and $\beta_i^{(3)}$ are independent, and follow a normal distribution with mean $\mu_\beta$ and variances equal to $\sigma_\beta$. In Scenarios 1-9 and 11-12, $\sigma_\beta$ is equal to zero and the parameters $\beta_i$ are identical across datasets (i.e., $\beta_1=\ldots = \beta_I$), while in Scenario 10 $\sigma_\beta>0$ and each study has different parameters $\beta_i$.

We note that a key assumption, which we call the \textit{equal conditional distributions} (ECD) assumption, has a central role in several borrowing methods. We say that the ECD assumption holds if
\begin{equation} \label{ecd_assumption}
    \begin{split}
        P(Y_{1j} | T_{1j} &= 0, \bm{X}_{1j} = \bm{x}) = P(Y_{2j} | T_{2j} = 0, \bm{X}_{2j} = \bm{x}) \\
        &= \dots = P(Y_{Ij} | T_{Ij} = 0, \bm{X}_{Ij} = \bm{x}),
    \end{split}
\end{equation}
for all pre-treatment patient profiles $\bm{x}$. In other words, the study-specific conditional distributions of the outcome, given the individual  pre-treatment profile, are identical. In our simulations we consider several violations of the ECD assumption, including unmeasured confounding (Scenario 9) and heterogeneous intercepts $\delta_i$ (Scenarios 6, 7, 11). The presence of heterogeneous intercepts $\delta_i$ can be due to factors that differ across data sources, for example measurement standards that vary across studies.

\subsubsection{Simulation Scenarios}\label{sec:scenarios}

The 12 scenarios that we consider are summarized in Table \ref{tab:scenarios}. 
Scenarios are specified by varying nine characteristics (features A to I in Table \ref{tab:scenarios}).
In Scenarios 1 to 4 it is ideal to leverage external data. All the external datasets have
identical covariate distributions $f_i$ 
as in the RCT (feature E, Table \ref{tab:scenarios}),
identical intercepts $\delta_1=\delta_2=\ldots = \delta_I$ in model \ref{eq:dgm} 
(feature F, Table \ref{tab:scenarios}),
the same regression parameters $\beta_1=\beta_2=\ldots = \beta_I$ (feature H, Table \ref{tab:scenarios}), and the outcome does not depend on 
the unmeasured covariate $X^{(3)}$ (feature G, Table \ref{tab:scenarios}).
Scenario 1 considers an RCT with sample size 100, and three external datasets of size 25. 
Patients in the RCT are randomized to experimental and SOC treatments in a ratio of 2:1. 
Scenario 2 considers a larger trial with a sample size  of 120 patients, 2:1 randomization, and one external dataset of size 30. 
Scenario 3 considers an RCT of size 80, 2:1 randomization, and seven external datasets of size 20. Scenario 4 considers an RCT of size 100, 1:1 randomization, and 3 external datasets of size 25.

Scenarios 5 to 12 are similar to Scenario 1, but introduce relevant differences between the RCT and the external datasets.
In Scenarios 5, 7 and 9, the distributions of pretreatment covariates $f_i$ differ across datasets (feature E, Table \ref{tab:scenarios}). 
In Scenarios 6, 7 and 11, the values of the intercepts $\delta_i$ in model \ref{eq:dgm} varies across datasets (feature F, Table \ref{tab:scenarios}). 
In Scenarios 6 and 7 the $\delta_i$'s have a nearly symmetrical distribution. In Scenario 11 the $\delta_i$'s have a markedly skewed distribution that differ from the normal distribution of the random effects in subsection \ref{sec:RE} to \ref{sec:PSC}.  
In Scenarios 8 and 9, the outcome depends on the unmeasured  covariate $X^{(3)}$ (feature G, Table \ref{tab:scenarios}).
In Scenario 10, the parameters $\beta_i$ used to generate the outcomes from model \ref{eq:dgm} differ across datasets (feature H, Table \ref{tab:scenarios}; $\sigma_\beta\neq 0$ in model \ref{eq:dgm}). 
Several methods discussed in Section \ref{sec:Methods} assume a logistic link function without  interactions between the pretreatment covariates $X^{(\cdot)}$.
Scenario 12  violates this assumption. 
The data-generating model of Scenario 12 is described in Table \ref{tab:scenario12}

\begin{table}[]
	\centering
	\begin{tabular}{cccc}
		\hline
		$x^{(1)}_{ij}$ & $x^{(2)}_{ij}$ & $T_{ij}$ & $P(Y_{ij}=1\mid \bm X_{ij}=\bm x_{ij},T_{ij}=t_{ij})$ \\
		\hline
		0 & 0 & 0 & 0.45 \\
		\hline
		0 & 0 & 1 & 0.71 \\
		\hline
		0 & 1 & 0 & 0.25 \\
		\hline
		0 & 1 & 1 & 0.51 \\
		\hline
		1 & 0 & 0 & 0.60 \\
		\hline
		1 & 0 & 1 & 0.86 \\
		\hline
		1 & 1 & 0 & 0.35 \\
		\hline
		1 & 1 & 1 & 0.61 \\
		\hline
	\end{tabular}
	\caption{The data-generating model of Scenario 12. 
	}
	\label{tab:scenario12}
\end{table}

We provide supplementary R code and an RMarkdown file that enables the reader to reproduce our simulation study \cite{kotecha_supplement_code}. The code generates the datasets $\{D_i \}_1^I$ and implements the methods outlined in Section \ref{sec:Methods}. 

\subsubsection{Operating Characteristics} \label{sec:OCs}

For each scenario and method, we simulated $S = 100,000$ independent replicates of the trial data $D_1$ and the external data collection $(D_2, \dots, D_I)$. We evaluated the described methods for hypothesis testing ($H_0: \tau \leq 0$) and estimating the treatment effects.
We focus on the following OCs, and report averages across $S$ simulations: the type 1 error rate, power, the bias $\frac{1}{S}\sum_{s}\left[\tau^{(s)}-\hat\tau^{(s)}\right]$ of the treatment effect estimates $\hat\tau$, where $\tau^{(s)}$ and $\hat \tau^{(s)}$ indicate the unknown and estimated treatment effect in simulation $s$, and (iv) the root mean squared error (rMSE) $\sqrt{\frac{1}{S}\sum_{s}\left[\left(\tau^{(s)}-\hat\tau^{(s)}\right)^2\right]}$ of the treatment effect estimates.

\begin{figure*}	
	\centering
	\includegraphics[width=16cm]{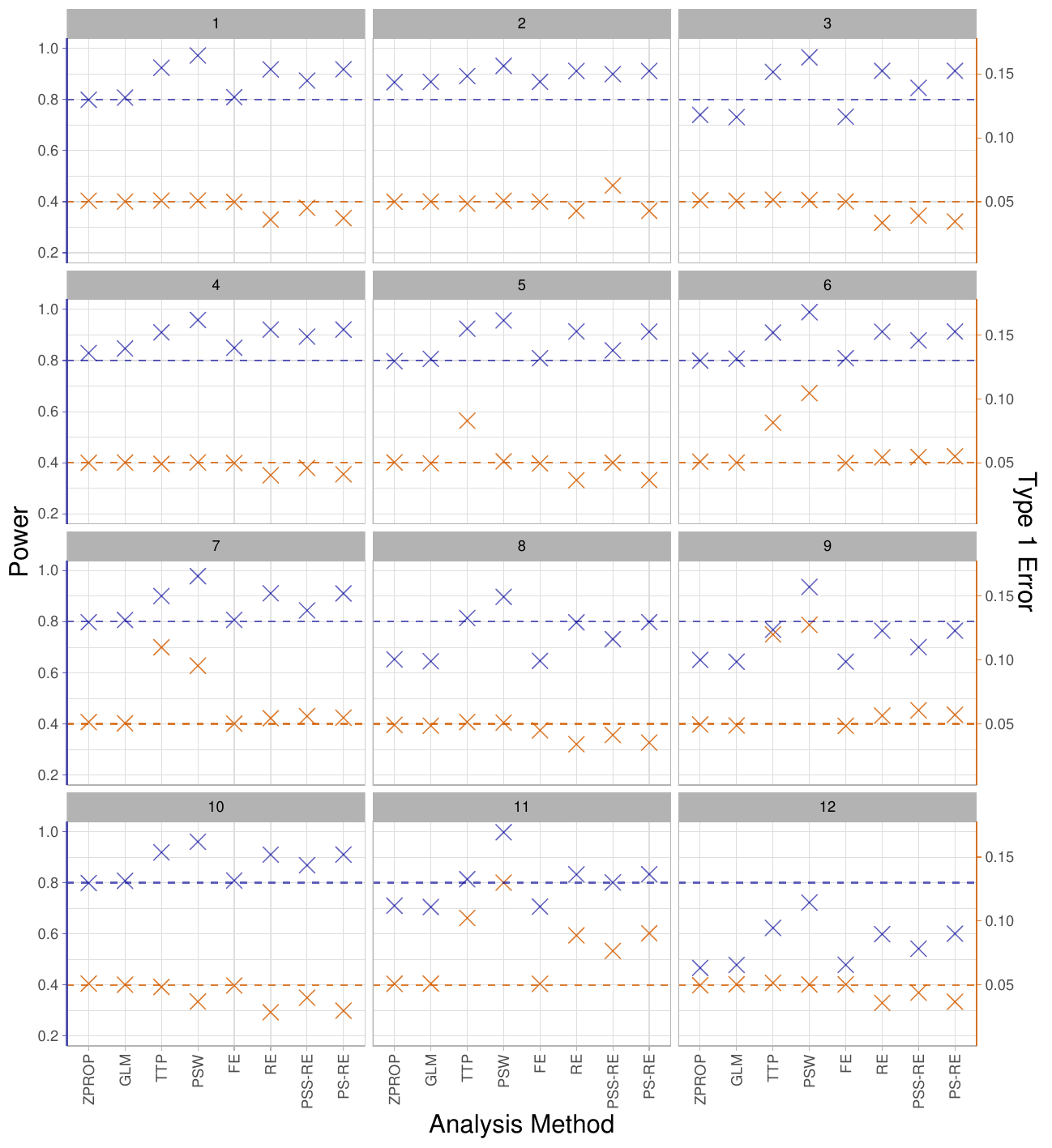}
	\caption{The methods for analyzing an RCT with external data described in Section \ref{sec:Methods} were compared across the scenarios described in Section \ref{sec:scenarios}. 100,000 simulations were carried out, with and without a treatment effect for each of 12 scenarios. Each of the 12 panels report results for one scenario. We report the Power (Blue, Left Axis) and Type 1 Error (Red, Right Axis) of each method. ZPROP: two sample Z test of proportions (no external data used). GLM: Logistic GLM (no external data used). TTP: Test-then-pool. PSW: Propensity score weighting. FE: Fixed effects model. RE: Random effects model. PSS-RE: Propensity Score Stratification and random effects. PS-RE: Propensity scores within random effects models.}
	\label{Fig:Power1}
\end{figure*}

\begin{figure*}
	\centering
	\includegraphics[width=16cm]{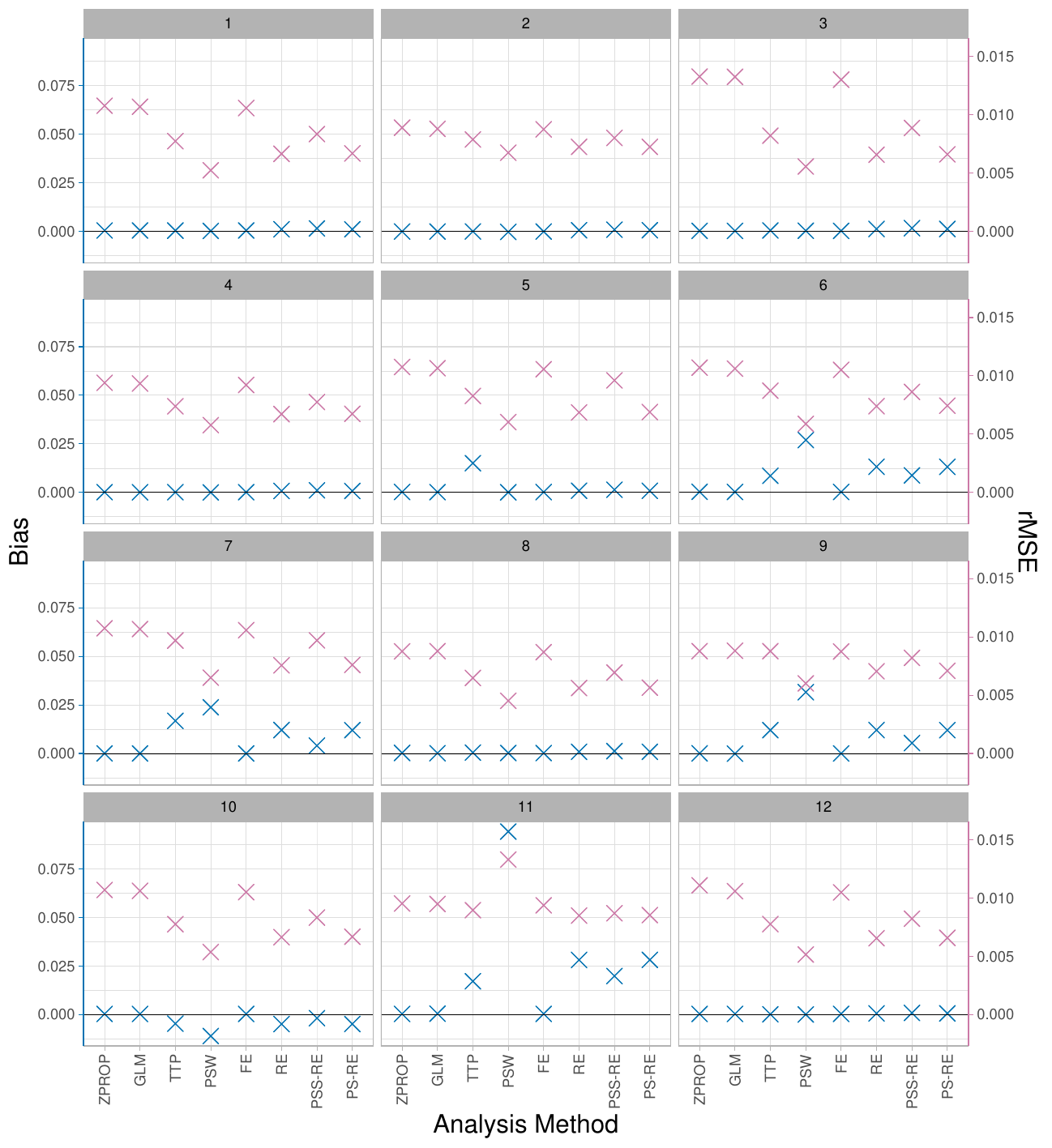}
	\caption{The methods for analyzing an RCT with external data described in Section \ref{sec:Methods} were compared across the scenarios described in Section \ref{sec:scenarios}. 100,000 simulations were carried out, with and without a treatment effect for each of 12 scenarios. Each of the 12 panels report results for one scenario. We report the Bias (Blue, Left Axis) and rMSE (Pink, Right Axis) of each method. ZPROP: two sample Z test of proportions (no external data used). GLM: Logistic GLM (no external data used). TTP: Test-then-pool. PSW: Propensity score weighting. FE: Fixed effects model. RE: Random effects model. PSS-RE: Propensity Score Stratification and random effects. PS-RE: Propensity scores within random effects models.}
	\label{Fig:BiasMSE}
\end{figure*}

\subsubsection{Simulation Results}\label{sec:results}

We compare the methods described in Section \ref{sec:Methods}. 
ZPROP and GLM use only the RCT dataset $D_1$ and ignore external data $D_2,\ldots, D_I$. 
As expected, both methods have a type I error rate close to the desired 5$\%$ level, and approximately equal power. 
In all scenarios with positive experimental treatment effects, ZPROP and GLM  as expected tend to have a lower power than the approaches that leverage external data. For example, in Scenario 6, ZPROP and GLM have 80\% and 81\% power, while all other methods have $\geq 88\%$ power (Figure \ref{Fig:Power1}, Panel 6, Blue). 
Both ZPROP and GLM provide nearly unbiased treatment effect estimates across all scenarios (Figure \ref{Fig:BiasMSE}, Blue), but they also have the highest rMSE (except in scenario 11). For example, in Scenario 1, the rMSE of ZPROP and GLM is 0.011 (Figure \ref{Fig:BiasMSE}, Panel 1), compared to a rMSE of $\leq 0.008$ for all alternative methods (except FE).

The TTP method includes or excludes external datasets $D_2,\ldots, D_I$ based on differences between response rates, without covariate adjustments. The type 1 error rate of the TTP method deviates from the target $\alpha$-level in all scenarios where the distribution of pretreatment variables varies across RCTs (Scenario 5, Figure \ref{Fig:Power1}, Panel 5, Red) and when $\delta_i$ parameters vary across RCTs (Scenario 6, Figure \ref{Fig:Power1}, Panel 6, Red). For instance, we observe a  type I error inflation of 12\% in Scenario 7 (Figure \ref{Fig:Power1}, Panel 7, Red). 
In Scenarios (5-7 and 9-11) where TTP exhibits substantial type I error rate inflations, we also observe substantially biased treatment effect estimates. For example, in Scenario 5, TTP has a bias of 0.015 (Figure \ref{Fig:BiasMSE}, Panel 5, Pink), compared to $\approx 0$ for ZPROP. 

The PSW method as expected has substantially inflated type I error rates in scenarios where the intercepts $\delta_i$ vary between datasets (Scenarios 6, 7 and 11). For example, in Scenario 6 the type I error rate is 0.11 (Figure \ref{Fig:Power1}, Panel 6, Red). 
In Scenario 9, where the outcome distribution depends on the unmeasured covariate, and the prevalence of the pre-treatment variables varies across studies, PSW has a type 1 error rate of 13\%. 
In Scenarios 6, 7, 9 and 11, where PSW exhibits inflated type I error rates, treatment effect estimates are also biased. For example, in Scenario 6, the bias is 0.027 (Figure \ref{Fig:BiasMSE}, Panel 6, Blue). In Scenarios 1-10 and 12, PSW has the lowest rMSE of all the methods. 
In Scenario 11, where the intercepts $\delta_i$ vary between RCTs, PSW does not account for differences between datasets. As a result, in Scenario 11, PSW has an rMSE of 0.013, the highest rMSE of all the methods (Figure \ref{Fig:BiasMSE}, Panel 11, Pink). 

The FE method described in Section \ref{sec:FE} maintains a type I error rate close to the target 0.05 throughout all scenarios (Figure \ref{Fig:Power1}, Red). The power of FE is similar to the one of GLM and ZPROP. For instance, in Scenario 1, both FE and GLM have a power of 0.81 while ZPROP has a power of 0.80 (Figure \ref{Fig:Power1}, Panel 1, Blue). FE exhibits near unbiased treatment effect estimates in all scenarios (Figure \ref{Fig:BiasMSE}, Blue). Across scenarios, the rMSE is similar to the rMSE attained by GLM and ZPROP. For example, in Scenario 1, FE, ZPROP, and GLM all have an rMSE of around 0.011 (Figure \ref{Fig:BiasMSE}, Panel 1, Pink). 

The RE and PS-RE methods described in Sections \ref{sec:RE} and \ref{sec:PSC} both use a random effects model. These two methods perform similarly throughout all scenarios. In Scenarios 1-10 and 12, both methods have a type I error rate of $\leq $0.057 (Figure \ref{Fig:Power1}, Red). Moreover RE and PS-RE have high power compared to the remaining methods, e.g. 0.91 for RE and PS-RE in scenario 2 compared to 0.8 for ZPROP, respectively (Figure \ref{Fig:Power1}, Panel 2, Blue). RE and PS-RE exhibit only minor bias (between $-0.00125$ and $0.00125$) in Scenarios 6-7 and 9-10 (Figure \ref{Fig:BiasMSE}, Blue) when the parameters $\delta_i$, $f_i$ or $\beta_i$ vary across datasets.
RE and PS-RE exhibit the second  rMSE (ranking from lowest to highest) of  the  
candidate methods in Scenarios 1-10 and 12, e.g. in Scenario 8, both have an rMSE of 0.006, compared to PSW's rMSE of 0.005 and ZPROP's rMSE of 0.009 (Figure \ref{Fig:BiasMSE}, Panel 8, Pink). 
In Scenario 11, the $\delta_i$ parameters  have a skewed distribution,
which deviates from the random effects' distribution assumed by the RE and PS-RE procedures. 
In this scenario, RE and PS-RE have a type I error rate of 0.09 (Figure \ref{Fig:Power1}, Panel 11, Red), substantially higher than the $\alpha=0.05$ target. In Scenario 11, RE and PS-RE  exhibit also substantial bias (Bias = 0.028, Figure \ref{Fig:BiasMSE}, Panel 11, Blue).

The PSS-RE and RE methods perform similarly, with some differences: 
PSS-RE has a type I error rate $\leq 0.06$ in Scenarios 1, 3-10 and 12 (Figure \ref{Fig:Power1}, Red). In Scenario 2 there is only one external dataset $D_2$, and PSS-RE has a type I error rate of 6.3\% (Figure \ref{Fig:Power1}, Panel 2, Red). 
%
%
In scenario 11, with a skewed distribution of  the $\delta_i$ parameters across datasets,
PSS-RE shows, similar to RE, an inflated type I error rate (7.7\%). 
PSS-RE has a higher power than methods that don't leverage external data (ZPROP and GLM), but (in most scenarios) has a lower power than the RE method. For instance, in Scenario 10, the PSS-RE, ZPROP and RE have a power of 87\%, 80\% and 91\%, respectively (Figure \ref{Fig:Power1}, Panel 10, Blue). In Scenarios 6, 7, 9, 10 and 11 (Figure \ref{Fig:BiasMSE}, Blue), where the parameters $\delta_i$, $f_i$ or $\beta_i$ vary across datasets, PSS-REs' treatment effects estimates exhibit substantial bias. 
Similarly, in all scenarios, the rMSE of PSS-RE is lower than for methods that don't leverage external data (ZPROP and GLM).

\begin{table*}[ht]

\SetTblrInner{rowsep=6pt}
 
\begin{tblr}{X[c,m,0.12\linewidth]X[c,m,0.22\linewidth]X[c,m,0.28\linewidth]X[c,m,0.28\linewidth]}
\toprule
\textbf{Scenarios} & \textbf{Description} & \textbf{Methods with superior operating characteristics} & \textbf{Methods with inferior operating characteristics} \\ \toprule
{1-5, 8, 10, 12} & {ECD holds} & {PSW} & {ZPROP, GLM, FE \\ TTP} \\ \midrule
{6, 7, 9} & {ECD violated \\ \& \\ RE model reasonable} & {RE, PSS-RE, PS-RE} & {ZPROP, GLM, FE \\ TTP, PSW} \\ \midrule
{11} & {ECD violated \\ \& \\  RE model assumptions violated} & {ZPROP, GLM, FE} & {RE, PSS-RE, PS-RE \\ TTP, PSW} \\ 
\bottomrule
\end{tblr}

\caption{Summary of the simulation results. 
We identified three groups of simulation scenarios. 
The third column, ``Methods with superior operating characteristics,'' identifies the methods that achieved both (a) approximately nominal control of false positives and (b) power greater than or (approximately) equal to the power of the other methods. The fourth column,
``Methods with inferior operating characteristics,'' identifies the methods that had either (a) high rates of false positives or (b) lower power (by at least 5\%) than any of the method that effectively controlled the false positive rate. The ECD is defined by equation \eqref{ecd_assumption}.}

\label{summary_table}
        
\end{table*}

Table \ref{summary_table} summarizes  our simulation results. It identifies three groups of simulation scenarios in which different methods to leverage external data present better OCs compared to the other candidate methods: 
\begin{enumerate}[noitemsep, topsep=0pt, parsep=6pt, partopsep=0pt]
    \item When the ECD assumption holds (Scenarios 1-5, 8, 10, and 12), PSW is the most powerful method and the treatment effect estimates are nearly unbiased. The RE methods are slightly less powerful but overall the OCs of the RE and PSW procedures are similar.

    \item When the ECD assumption is violated and the differences  between the study-specific outcome distributions conditional on the patient pre-treatment profiles are well described by a model with random intercepts (Scenarios 6, 7, and 9), RE, PSS-RE, and PS-RE present minimal bias and  higher power compared to data analyses without external controls. These methods account for potential differences between the study-specific   outcome distributions  conditional on the patient pre-treatment profile.
            
    \item In Scenario 11 the ECD assumption is violated and the differences between the study-specific  outcome distributions conditional  on the individual pre-treatment profiles are \textit{not}  adequately captured by a model with Gaussian random intercepts. In this case FE, ZPROP, and GLM yield better OCs relative to the other candidate methods. We note that with  FE the external data have limited influence on the treatment effect estimate (unless the covariates are highly predictive of the outcomes), while with ZPROP and GLM the external data are not used.
\end{enumerate}
These results are aligned with the causal inference literature, as well as the assumptions and construction of  each method.

The relative performance of these methods, compared to methods that exclude  external data, share a similarity with the so-called Hodges' estimator paradox \citep[Chapter 8.1]{van_der_vaart_asymptotic_2007}. Consider a sequence of estimators $\hat{\theta}_n$ with asymptotic distribution $\sqrt{n} \left( \hat{\theta}_n - \theta \right) \overset{d}{\to} L_\theta$, where $L_\theta$ is  indexed by a real parameter $\theta$. This sequence can be compared to Hodges' estimator, which is defined as 
\begin{align*}
    \hat{\theta}^H_n = \begin{cases}
        \hat{\theta}_n, & \text{if } |\hat{\theta}_n| \geq n^{-1/4}, \\
    0, & \text{if } |\hat{\theta}_n| < n^{-1/4}.
    \end{cases}
\end{align*}
When we compare finite sample OCs like the mean squared error, Hodges' estimator $\hat{\theta}^H_n$ tends to (i) perform better than $\hat{\theta}_n$ if $\theta$ is equal or very close to 0, (ii) perform nearly identically to $\hat{\theta}_n$ for $\theta$ far from 0, and perform worse than $\hat{\theta}_n$ for intermediate values of $\theta$ \citep[Figure 8.1]{van_der_vaart_asymptotic_2007}. Similarly, compared to not using external data, several methods that integrate external data present OCs that are superior to analyses based only on RCT data when there are few or no distortions and the method-specific assumptions hold. Also, the OCs for several methods are nearly identical if  there are  large distortions that make their detection trivial (e.g., PSS-RE and ZPROP in Scenario 9). Finally, as illustrated in our simulation study, in intermediate scenarios between these two extremes, methods using external data can have worse OCs than analyses based only on RCT data (e.g, PSS-RE and ZPROP in Scenario 11).

These results highlight, as expected, that  there is no effective borrowing method that completely avoids biased treatment effects estimates across \textit{all} potential scenarios. Our comparisons did not identify a single statistical procedure that consistently outperforms all the others in every scenario. While our 12 simulation scenarios ---defined by the nine features in Table \ref{tab:scenarios}--- cover a broad range of clinical trial contexts, the space of possible scenarios is much larger and includes other potential distortion mechanisms that we did not consider. For example, we did not consider scenarios in which there are slight variations of the definition of the disease over time \citep[see for example][]{melhem_updates_2022}. Our results show that the strengths and limitations of different methods for incorporating external data depend on trial-specific factors such as sample size, the number and heterogeneity of external data sets, and the level of standardization in recruitment and outcome assessment across data sources. As such, decisions about whether to leverage external data --- and how to do so --- need to be context-specific.

Indeed, Figure \ref{Fig:Power1}, Figure \ref{Fig:BiasMSE}, and Table \ref{summary_table} show that the methods differ substantially in (a) the magnitude of performance gains (e.g., power or MSE, compared to analyses based only on RCT data) under ideal scenarios, (b) the set of scenarios (RCT and ED data distributions) that satisfy their assumptions, and (c) their robustness when these assumptions are violated. These differences underscore the importance of selecting an appropriate method when external data are available. For example, consider two similar RCTs that plan to incorporate external data. In the first, outcome assessments are based on the same procedure in the trial and external data, reducing the risk of bias. In the second, outcome assessments used different procedures in the trial and external data, increasing the risk of bias. Our simulations suggest that PSW may be suitable for the first trial but should be avoided for the second. As we show in the following section, investigators can use tailored simulation scenarios to explore plausible distortion mechanisms and guide decisions about whether and how to integrate external data.

\subsection{Model-free evaluation of the methods using a GBM data collection}\label{Sec:Application}

We compare  the methods described in Section \ref{sec:Methods} in retrospective analyses of a collection of individual patient-level GBM datasets (IPLD). 
The data collection was described in \cite{rahman_accessible_2023}.
The objective is to provide specific recommendations for the design and analysis of future GBM trials. 
We use a resampling procedure, similar to the algorithms described in \cite{ventz_design_2019}, to generate {\it in silico} trials and provide estimates of OCs of trial designs, including type I error rate, power, bias, and the rMSE. Investigators who plan to use external data to analyze RCTs in other disease areas can conduct similar simulations, tailored to the specific clinical setting. Other validation approaches for a critical assessment of study designs and data analysis plans, different from our resampling approach, have been previously discussed in the literature; see for example \cite{shadish_can_2008}.

{\it Data.} The analyses are based on  four GBM datasets, 
a phase 3 clinical study \cite{chinot_bevacizumab_2014} (NCT00943826) with 458 patients, 
two phase 2 trials (PMID: 22120301 and NCT00441142) with 16 and 29 patients, respectively \cite{cho_adjuvant_2012, lee_multicenter_2015}, and a real-world dataset with 663 patients from the Dana-Farber Cancer Institute (DFCI). 
We use data from patients treated with temozolomide and radiation therapy (TMZ+RT), the current SOC in GBM. 
Pretreatment characteristics included age, sex, Karnofsky performance status, methylation status, and the extent of tumor resection. We consider a binary outcome, overall survival 12 months after treatment initiation (OS-12). 

{\it Generating in silico RCTs.}
Our algorithm samples at random, without replacement, a subset of patients from the control arm of one trial \cite{chinot_bevacizumab_2014} to produce the in silico RCT dataset $D_1$. 
For each sample size $n_1\in \{75, 80, \ldots, 175\}$, we repeated the following four steps 10,000 times: 

(i) We randomly select, without replacement, 
$n_1$ 
individual-level data points (pre-treatment profiles and outcomes) 
from the TMZ-RT control arm of the \cite{chinot_bevacizumab_2014} trial. 

(ii) We then randomize these $n_1$ patients in a ratio r:1 to experimental and control arms of the {\it in silico} RCT, to obtain a dataset $\tilde D_1$. 
The joint distributions of outcomes and pretreatment variables in the control and experimental arms of the \textit{in silico} trial are identical and match the distribution in the control arm in the actual \cite{chinot_bevacizumab_2014} trial. 

\begin{landscape}
    \begin{figure}[htbp!]
        \centering
	\includegraphics[width=24cm]{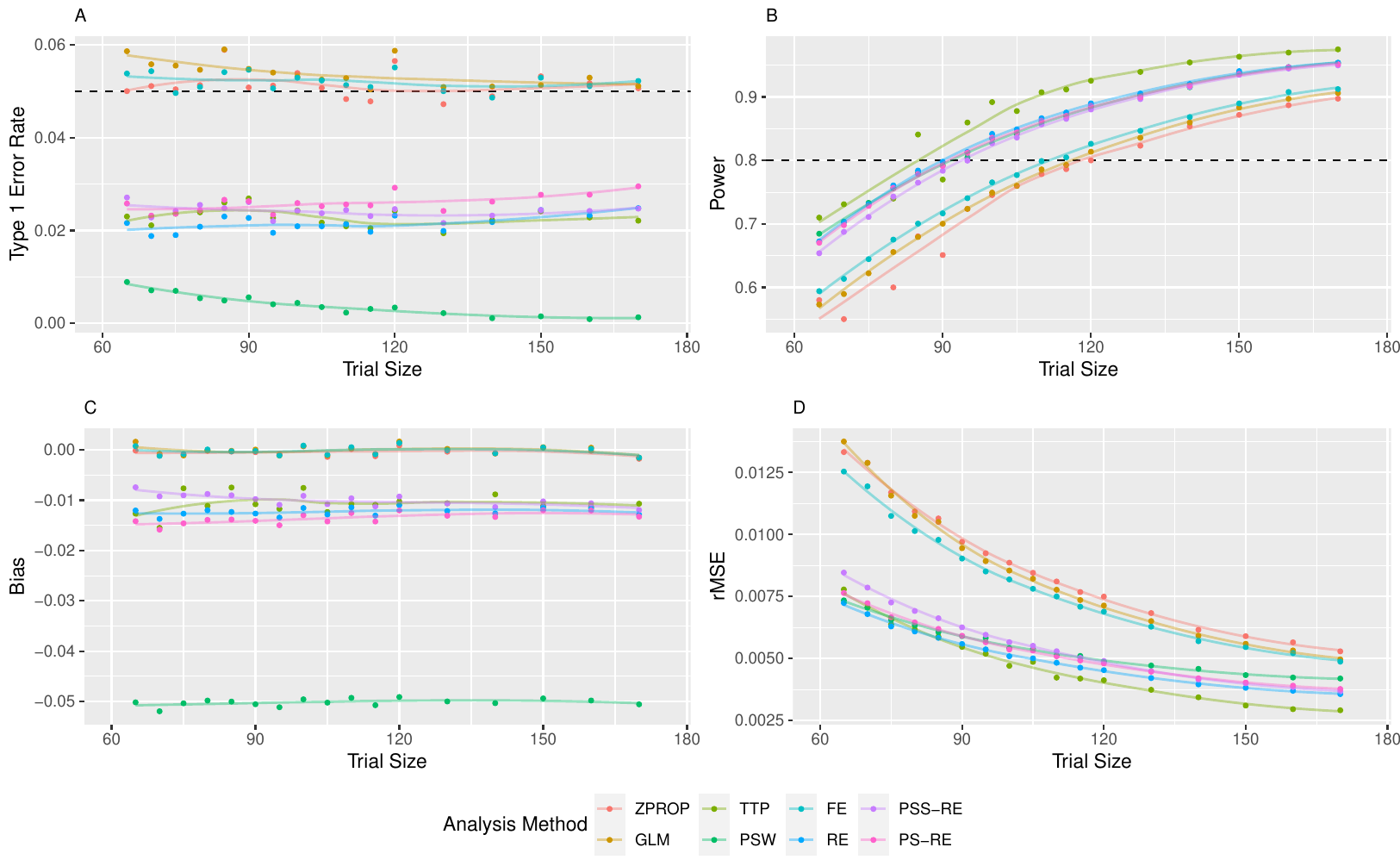}
	\caption{ We sampled individual data points (outcomes and pretreatment covariates) from completed clinical trials and observational studies in GBM in order to mimic an RCT. For each trial size (X-axis), 20,000 simulations were run, half with no treatment effect and half with a positive treatment effect. The Y axes display the OCs for  8 analysis methods. ZPROP: 2 sample Z-test for proportions (no external data used). GLM: Logistic GLM (no external data used). TTP: Test-then-pool. FE: Fixed Effects model. RE: Random Effects model. PSW: Propensity score weighting. PSS-RE: Propensity Score Stratification with Random Effects Model. PS-RE: Propensity Score Covariate Adjustment with Random Effects. In subfigures A and B the dashed horizontal line shows the target Power of 0.8 and the target T1E of 0.05 respectively.}
	\label{fig:rdpt}
    \end{figure}
\end{landscape}

(iii) We use the IPLD of patients treated with TMZ+RT in the remaining datasets as external data sources $D_2$, $D_3$ and $D_4$, respectively. 
These datasets are fixed across all 10,000 iterations 
($n_2=16, n_3=29$ and $n_4=663$).

(iv) We apply each of the methods in Section \ref{sec:Methods} to 
provide inference on treatment effects based on the 
datasets $\mathcal{D}=\{\tilde D_1, D_2, D_3, D_4\}$. 

Steps (i-ii) of the algorithm generate \textit{in silico} RCTs with null treatment effects. 
When we needed to simulate \textit{in silico} RCTs with positive treatment effects, we slightly modified the algorithm by adding an additional component to step (ii). 
For each patient $j$ in the {\it in silico} experimental arm of the RCT with negative treatment response ($Y_{1j}=0$), we randomly generated a binary random variable $R_j$, with success probability $P(R_j=1)=0.2$, and set $Y_{1j}=R_j$. That is, in our {\it in silico} experiments $P(R_j=1)$ represents the effects of the experimental 
treatment for patients who would have experienced a negative outcome under the control therapy.

{\it Results.}
Figure \ref{fig:rdpt} shows the estimated type I error rate, power, bias, and the rMSE for all eight methods for various trial sizes $n_1$ (x-axis). Figure \ref{fig:rdpt}(A) shows a type 1 error rate of $\leq $0.06 for all methods and for all the trial sizes between 75 and 175. 
The methods that use external data, with the exception of FE, have type I error rates below 0.035. 
Figure \ref{fig:rdpt}(B) shows that the methods that leverage external data, with the exception of FE, require $\leq  95$ patients to attain a power of 80\%, 
whereas approaches that do not use external data (ZPROP and GLM) required more than 115 patients to achieve a power $\ge 80\%$. In other words, using external data in the analysis of GBM RCTs  can 
reduce  the sample size (by approximately 20$\%$) necessary to achieve the 
80\% power.
This shortened the average study duration by approximately 12$\%$ in our analyses, where we 
assumed an average enrollment rate of 6 patients per month.

Figure \ref{fig:rdpt}(C) 
shows that the methods that do not use external data (ZPROP and GLM) and the FE are nearly unbiased, whereas 
in our retrospective analysis
the other methods that leverage external data exhibit a  negative bias (-0.05 TTP, $\approx -0.01$ for PSW, RE, PSS-RE and PS-RE). 
Figure \ref{fig:rdpt}(D) shows how  methods that use external data (TTP, PSW, RE, PSS-RE, PS-RE) exhibit a lower rMSE than the others (ZPROP and GLM) for sample sizes $n_1$ between 50 and 200. 

RE appears to be an attractive choice for future RCTs in GBM that use external data, 
based on favorable performances in simulations (Section \ref{sec:results}) and our 
retrospective analysis in GBM (Figure \ref{fig:rdpt}). In our subsampling experiments,
RE-based methods (RE, PSS-RE and PS-RE) exhibited power improvements (between 5$\%$ and 15$\%$) compared to methods that do not use external data (ZPROP and GLM) and maintained a type I error rate $\leq 3\%$ (Figure \ref{fig:rdpt}). 

Similarly, in our simulation study (\ref{Fig:Power1}), RE had a power above $75\%$, higher than ZPROP and GLM,  
with sample sizes of $n_1 \geq 100$ (Figure \ref{Fig:Power1}, Blue) and controlled the type I error rate below 6$\%$, except for the Scenario in which a high type I error rate was expected (Scenario 11, Figure \ref{Fig:Power1}, Panel 11, Red).

Although the TTP and PSW methods performed well in our analyses in GBM, we do not recommend their use due to the high risks of type I error rate inflations (Figure \ref{Fig:Power1}). 

We evaluated if RE's control of the type I error rate 
in our retrospective analyses
was  driven by the small negative bias seen in Figure \ref{fig:rdpt}(C) or not.
Patients in the observational dataset from DFCI, which was used as $D_4$, tend to have slightly better outcomes than those in \cite{chinot_bevacizumab_2014} (which was use to generate the RCTs) conditionally on age, gender, Karnofsky performance status, and the resection status. 

We therefore repeated Steps i-iv of our resampling algorithm, swapping the two largest datasets so that the RCT data ${D}_1$ was generated by sampling from the real-world DFCI dataset and $D_4$ was the entire control arm of \cite{chinot_bevacizumab_2014}. 
We ran 3000 simulations with $n_1=120$ and found that as expected  RE exhibited a small positive bias of $+0.03$, the rMSE was 0.04, and type I error rate was maintained at $\leq 6 \%$.

\section{Discussion}

We considered the analysis of an RCT  augmented with patient-level data from previous clinical trials and real-world datasets. We compared six candidate procedures, focusing on popular  methods with available and easy-to-use software. We used simulations and completed GBM studies to compare the six statistical procedures. 

Our simulations included a broad set of scenarios, including ideal scenarios in which it is efficient and straightforward to leverage information from external datasets. We also explored scenarios with various   distortion mechanisms, such as unmeasured confounding and model misspecification. In these simulations, we showed that leveraging external data  through an  appropriate analysis method (e.g. PSW, RE, PSS-RE, or PS-RE)  can  increase the RCT power. Some of the methods that we considered, for example TTP, exhibited a high risk of false positive results and biased  estimates in the presence of  minimal confounding mechanisms or other study-to-study differences. Because of these risks of bias, procedures to integrate external data into RCT analyses should (a) incorporate adjustments for potential confounders and, ideally (b) account for potential study-to-study differences  between the  distributions of the outcome conditional on pre-treatment covariates (i.e., potential violations of the ECD assumption in equation \eqref{ecd_assumption}). Among the methods that we considered in our simulations, the TTP approach does not account for potential confounders and the PSW procedure does not account for potential violations of the ECD assumption.

Our data analyses focused on GBM; in this  disease  area there has been substantial  interest in leveraging prior trials and observational datasets to improve the design and analysis of future studies \cite{avalos-pacheco_validation_2023, rahman_accessible_2023}. We used a collection of patient-level GBM datasets to compare statistical methods. In our data collection, the variations of pretreatment profiles and outcome distributions across studies (see \cite{rahman_accessible_2023} for details) are representative of  discrepancies that might arise in the integration of EC data in future trials. We used a resampling algorithm that generates realistic joint distributions of patient pretreatment profiles  and  outcomes. Our results suggest that the use of random effects models (RE) can provide a suitable solution for future phase 2 GBM RCTs that leverage external data.

We evaluated several established methods with available off-the-shelf software, without the intention of comprehensively evaluating the broad range of methods to integrate external data in the analysis of RCTs. Several other methods can be used to incorporate external data, including Bayesian and machine learning techniques \cite{chandra_bayesian_2023, li_target_2020, jiang_elastic_2023, hobbs_hierarchical_2011}. Another notable approach that we did not examine is matching. Matching  procedures  pair each patient in the RCT experimental arm with one or more patients in the external data with similar pre-treatment covariates \citep{lin_matching_2023}. A variety of algorithms can be used to assess similarity,  match patients, and estimate treatment effects; see for example \cite{hansen_optimal_2006}, \cite{stuart_matching_2010}, and \cite{rosenbaum_modern_2020}.
    
Our data analysis is informative for planning future GBM trials, but similar analyses in other disease areas are likely to produce different results. In particular, in some disease settings relevant limitations (e.g., measurement errors) of the available datasets, from completed trials or electronic health records, might translate into poorer OCs of trials that leverage external datasets.

\textit{Outcome types}. 
For ease of exposition we considered only binary outcomes in this paper. Other types of outcomes are  frequently used in clinical studies, for example patient survival time. The methods that we discussed can be adapted to studies with non-binary outcomes. When the outcome is a continuous or count variable, adapting these methods is relatively simple. We can use the same definition of the treatment effect $\tau$ that we adopted, the difference of means \eqref{tx_eff_def}. For example, the TTP procedure can use an appropriate two-sample test to detect potential differences between the outcome distributions in the control arm of the RCT and the external control group, and the estimator $\hat{\tau}$ defined in Section 2.1. The FE and RE procedures can use different likelihood functions (e.g., a Poisson regression model) together with the same plug-in estimators of $\tau$ (equations \ref{fe_tau_hat} and \ref{re_tau_hat}). Similarly, the propensity score methods (PSW, PSS-RE, PS-RE) can be adapted for the analysis of continuous or count outcomes. When the outcome is a time-to-event variable, the treatment effect is typically defined by hazard ratios or differences between restricted mean survival times. In these cases the methods that we discussed can be applied using survival models \citep{aalen_event_2008, royston_restricted_2013, zhao_restricted_2016}.

\subsection{Practical guidelines} 

Some practical guidelines on the integration of external data into clinical trials emerged from our comparative analyses.
    
\textit{First, we recommend conducting tailored simulation studies to guide trial design and analysis method selection.} 
To choose an appropriate method for integrating external data, analysts should conduct simulations that are realistic and tailored to the specific clinical setting. As expected, we found no single method that consistently outperformed the others across all trial scenarios. For example, in our model-based simulations (Figure \ref{Fig:Power1} and Table \ref{summary_table}) the random effects methods (RE, PSS-RE, and PS-RE) had high power and controlled  the Type I error rate in most scenarios, but performed poorly when their assumptions were violated, such as when the study-specific intercepts deviated substantially from normality (Scenario 11). A similar trade-off between efficiency gains and the control of false positive results in trials incorporating external data has been discussed by \cite{koppschneider_power_2020}.
    
Tailored simulations can be based on  a data collection with several datasets in the same disease setting, as we did in our model-free GBM simulations (Section 3.2). Researchers can generate in silico trials by resampling these datasets \citep{rahman_accessible_2023}, with the option of adding perturbations such as measurement error to introduce additional distortions. These perturbations help evaluate the robustness of the planned analyses. Disease-specific data collections are becoming increasingly available. The GBM data collection that we used includes more than 1,200 patients treated with the SOC, with individual pretreatment covariates and outcomes from six completed RCTs and two institutional databases. This resource is highly valuable for designing and analyzing future GBM studies. Similar  efforts in other disease areas include, for example, Project Datasphere \citep{green_project_2015}, Project YODA \citep{bierer_global_2016} and Vivli \citep{krumholz_yale_2016}. Simulations can also be tailored to a specific  disease  based on disease-specific meta-analyses and other studies that provide information about conditional or marginal outcome distributions across recent trials \citep[see for example][]{vanderbeek_randomize_2019}.
    
\textit{Second, the tuning parameters that regulate the influence of external data should be selected carefully.} Many borrowing methods include parameters that control how strongly the external data influence the final inference on treatment effects. Examples include the scaling factor $C$ in PSW and the penalty parameter $\eta$ in the RE procedure. These parameters can be pre-specified in the trial protocol based on simulations tailored to the trial design and clinical setting. Alternatively, they can be selected adaptively at trial completion based on the observed data \citep{lei_cross-validation_2020, bischl_hyperparameter_2023}. While we did not explore these adaptive procedures to select tuning parameters, we consider the adaptive approach  an important area for future research.
    
\textit{Third, we recommend  assessing  the robustness of the  trial results with respect to (a) plausible violations of assumptions on  the external controls and (b) alternative decisions about the available datasets that are used as external controls.} 
Incorporating non-randomized external controls into a trial can raise valid concerns about bias and the scientific rigor of reported findings. To address these concerns, we recommend planning sensitivity analyses and highlight two possible types. The first type assesses whether the trial's primary result (e.g., evidence of a positive treatment effect) would be contradicted if investigators were aware of violations of a key assumption about the external data, and accounted for such violations through statistical modeling. In particular, the framework of \citep{vanderweele_bias_2011} can  be used  to estimate if a primary result is robust    with respect to the influence of a hypothetical unmeasured confounder when this influence is  expressed by an odds-ratio or a similar metric. The second type of sensitivity analysis examines whether the primary findings are confirmed when we consider reasonable alternative decisions about the selection of the external datasets \citep{rahman_identifying_2025} and the statistical procedure to integrate external data. This can be done, for example, by re-analyzing the RCT including fewer external datasets or using different values of relevant tuning  parameters (e.g., smaller values of $C$ in  the PSW procedure). If sensitivity analyses confirm the primary results, they reduce  concerns that the primary results are merely caused by hypothetical artifacts and  distortion mechanisms in the external data.
    
\subsection{Potential innovations in the design and analysis of RCTs that leverage external data}

We also highlight potential directions of statistical innovation  to improve the integration of  external data in future RCTs.

First, there is a need for computational strategies to make the design of trials that integrate external data more efficient and easier. As we  discussed, the relative performance of different methods for data analyses --- measured by metrics such as power and bias --- vary considerably across scenarios. This implies that the selection of a design with an  appropriate data analysis plan depends on trial-specific factors, including the patient population as well as the number, size, and characteristics of the external datasets, such as the  measurement standards for prognostic variables and outcomes. The choice  of the data analysis plan should therefore be guided by both the statistical properties of  candidate procedures to integrate external data (e.g., the assumptions necessary for a rigorous control of false positives; \cite{koppschneider_power_2020}) and by tailored simulations. However,  simulations often require substantial time and computing resources, particularly when (a) many scenarios need to be explored (e.g., by varying sample sizes, prognostic factors' distributions, and treatment effects) and/or (b) the trial design includes computationally intensive steps, such as Bayesian interim decisions based on Markov chain Monte Carlo algorithms. Moreover, several designs and data analysis procedures involve tuning parameters (e.g., the $C$ parameter in PSW procedure) which are selected based on  simulations. Although interpolation strategies have been proposed to reduce the computing time  \citep[e.g.,][]{muller_optimal_1995, golchi_estimating_2022, golchi_estimating_2024}, further methodological developments are  needed. 
    
Second, there is a need to better characterize how the benefits (e.g., increased power) and risks (e.g., bias) of using external data vary across clinical settings. Interpretable metrics, like power improvements and estimation bias can be useful to identify contexts --- including diseases and phases of the drug development process  --- in which external data tend to be particularly beneficial. For example, we can compare (a) a small Phase 2 trial in a rare disease with a short median survival, and (b) a large Phase 3 registration trial in a more common disease with better prognosis and an experimental drug already approved for other conditions. The expected benefits and risks of integrating external data  differ significantly between these two contexts. These expectations depend on factors such as  regulatory requirements for drug approval \citep{wallach_us_2018}, treatment costs, and availability of high-quality standardized data \citep{sugden_patterns_2020}. However, the available resources to evaluate the risks and benefits of using external data --- such as data collections \citep[e.g.,][]{green_project_2015}, software and literature quantifying benefits and risks across disease settings --- are limited. Existing contributions often focuses on a single disease \citep{polley_leveraging_2024} or a narrow range of simulation scenarios. 

Third, there is a need for interpretable sensitivity analysis methods that assess the robustness of trial findings (e.g., a positive treatment effect) to violations of assumptions. While sensitivity analysis methods are well developed in the epidemiological and causal inference literatures \citep{cornfield_smoking_1959, rosenbaum_assessing_1983, imai_causal_2010, vanderweele_sensitivity_2017}, they have been developed for non randomized studies. For this reason, future statistical developments and the study of novel sensitivity analysis methods for randomized trials that incorporate external controls have the potential to improve investigators' assessment and reporting of the degree of robustness of RCT findings.

%
%

%

\begin{funding}
    We gratefully acknowledge support from the National Institutes of Health under grants R01LM013352 (LT and DES), T32CA009337 (DES and GS), and 5P30CA077598-25 (SV).
\end{funding}


\begin{supplement}
    \stitle{Supplement Simulation Results}
    \sdescription{Simulation results for alternative values of $C$ in the PSW method.}
\end{supplement}

\begin{supplement}
    \stitle{Supplementary R Code}
    \sdescription{Code used to implement each method and conduct simulations.}
\end{supplement}


\bibliographystyle{imsart-number} 

\bibliography{ec_sts_full.bib}
\end{document}


\maketitle

\renewcommand\thesection{S\arabic{section}}
\setcounter{section}{0}

\section{Simulation results for alternative $C$ in PSW}

\begin{figure}
        \centering
        \includegraphics[width=0.7\textwidth]{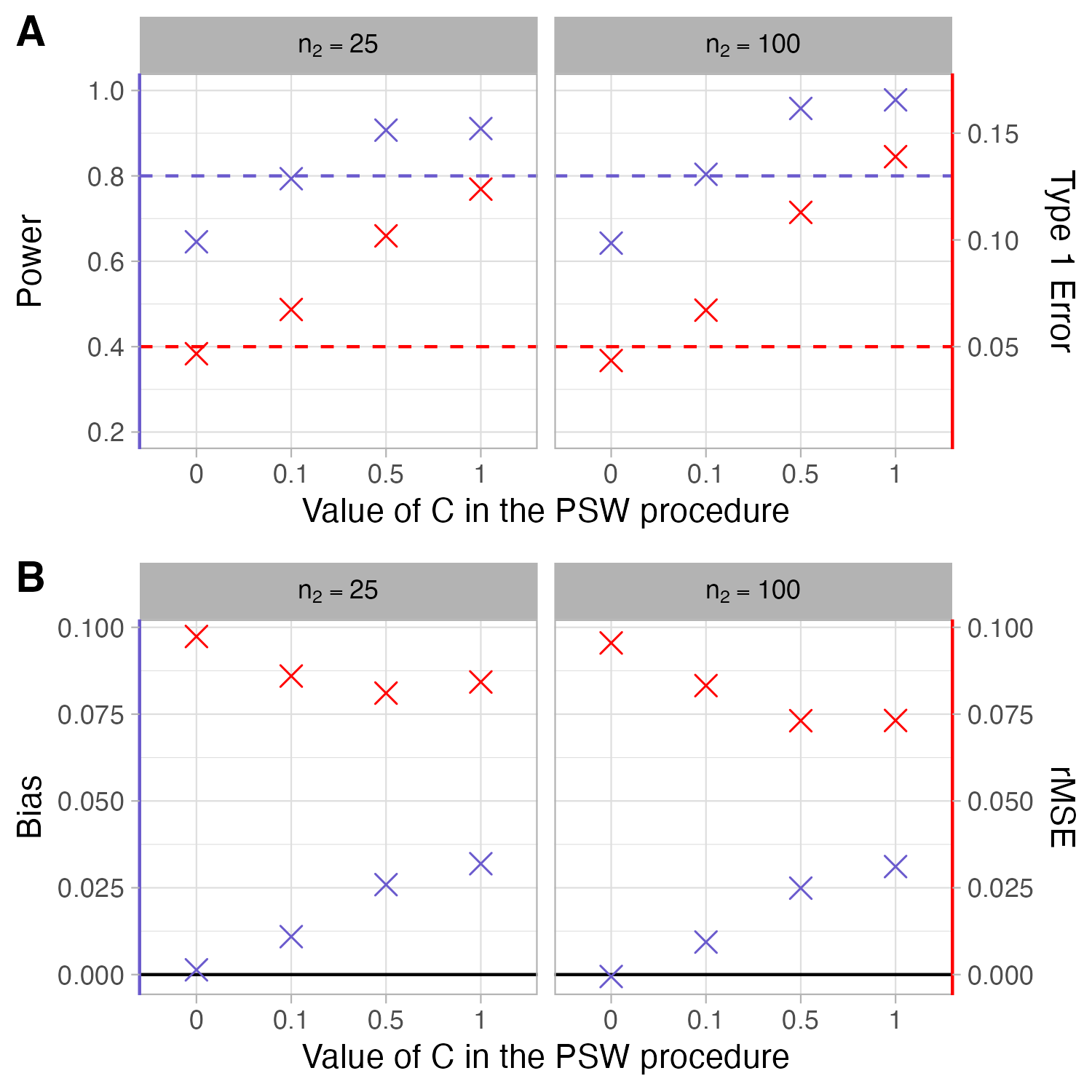}
        
        \caption{Comparison of PSW procedures with different values of the $C$ parameter. The comparisons are based on simulations, using Scenario 9 with external datasets of identical size $n_2 = n_3 = n_4 = $ 25 or 100. Blue denotes power (panel A) or bias (panel B) and red denotes type 1 error (panel A) or rMSE (panel B).}
        \label{c_fig}
    \end{figure}

Figure \ref{c_fig} (below) shows simulation results comparing the use of PSW with various values of the parameter $C$ to show how the bias, MSE, type I error, and power  depend on the choice of $C$. These simulations are based on Scenario 9 (with unmeasured confounding) and we varied the size of the external datasets. The results indicate that, as expected, the choice of $C$ influences major OCs. 

